\documentclass[12pt]{article}
\usepackage{amsfonts,amssymb,amsmath,mathrsfs}
\usepackage{hyperref}
\newcommand{\beq}{\begin{equation}}
\newcommand{\eeq}{\end{equation}}
\newcommand{\vp}{\vphantom}

\newcommand{\pt}{\partial}
\newcommand{\bs}{\boldsymbol}
\newcommand{\al}{\alpha}
\newcommand{\bt}{\beta}
\newcommand{\g}{\gamma}

\newcommand{\de}{\delta}

\newcommand{\la}{\lambda}
\newcommand{\La}{\Lambda}
\newcommand{\s}{\sigma}
\newcommand{\mc}{\mathcal}
\newcommand{\mr}{\mathrm}
\newcommand{\un}{\underline}

\newcommand{\coset}{PSU(2,2|4)/(SO(1,4)\times SO(5))}

\begin{document}
\begin{center}
{\Large\textbf{Supertwistor formulation for massless superparticle in $AdS_5\times S^5$ superspace}}\\[0.3cm]
{\large D.V.~Uvarov\footnote{E-mail: d\_uvarov@\,hotmail.com}}\\[0.2cm]
\textit{NSC Kharkov Institute of Physics and Technology,}\\ \textit{61108 Kharkov, Ukraine}\\[0.5cm]
\end{center}
\begin{abstract}
Starting with the first-order formulation of the massless superparticle model on the $AdS_5\times S^5$ superbackground and presenting the momentum components tangent to $AdS_5$ and $S^5$ subspaces as bilinear combinations of the constrained $SU(2)$-Majorana spinors allows to bring the superparticle's Lagrangian to the form quadratic in supertwistors. The $SU(2,2|4)$ supertwistors are assembled into a pair of $SU(2)$ doublets, one of which has even $SU(2,2)$ and odd $SU(4)$ components, while the other has odd $SU(2,2)$ and even $SU(4)$ components. They are subject to the first-class constraints that generate the $psu(2|2)\oplus u(1)$ gauge algebra. This justifies  previously proposed group-theoretic definition of the $AdS_5\times S^5$ supertwistors and allows to derive the incidence relations with the $(10|32)$ supercoordinates of the $AdS_5\times S^5$ superspace. Whenever superparticle moves within the $AdS_5$ subspace of the $AdS_5\times S^5$ space-time, twistor formulation of its Lagrangian involves just one $SU(2)$ doublet of $SU(2,2|4)$ supertwistors with even $SU(2,2)$ and odd $SU(4)$ components. If in addition particle's 5-momentum is null, four first-class constraints which are the $su(2)\oplus u(1)$ generators single out upon quantization the states of $D=5$ $N=8$ gauged supergravity multiplet in the superambitwistor formulation.
\end{abstract}
\setcounter{equation}{0}
\def\theequation{\thesection.\arabic{equation}}

\section{Introduction}

Incidence relations for the Penrose $SU(2,2)$ twistors \cite{Penrose} and Penrose-Ferber $SU(2,2|N)$ supertwistors \cite{Ferber}\footnote{For another possibility of the definition of (super)twistors for $D=4$ Minkowski (super)space see \cite{theta-twistor}.} that come out as solutions of the twistor equation and its superextensions can also be derived from the (super)twistor formulation of the massless particle model in 4-dimensional (super)space \cite{Ferber}, \cite{Shirafuji}, \cite{bbcl},  \cite{Eisenberg}, \cite{Plyushchay}, \cite{Gumenchuk}, \cite{DVVolkov}. (Super)twistor formulation is known to combine manifest and linearly realized (super)conformal symmetry with the simple and irreducible realization of the gauge symmetries. These features provided strong motivation to study (super)twistor formulations also for massive (super)particles \cite{Zima}, \cite{Bette}, \cite{Picon1}, \cite{dAIL}, \cite{Mezincescu'13}, null and tensile (super)strings \cite{Ilyenko}, \cite{U-CQG06}, \cite{BdAM}, \cite{FL06} and membranes \cite{FL07} in 4-dimensional Minkowski (super)space and in higher string-theoretic dimensions \cite{bc}, \cite{Berkovits'90}, \cite{Howe-West}, \cite{BLS}, \cite{BdAPV}, \cite{Picon2}, \cite{U-CQG07} \cite{Routh}, \cite{Bandos'14}.\footnote{Yet another approach to the construction of (super)twistors in higher-dimensions relies on the properties of pure spinors (see, e.g. \cite{Cherkis}, \cite{Berkovits'10}).} (Super)twistors also appear rather efficient in presenting scattering amplitudes of massless particles not limiting to 4-dimensional space-time (see, e.g. \cite{Elvang} and references therein).

Since $SU(2,2)$ is not only the covering of the conformal group of
4-dimensional Minkowski space-time but also is the isometry group
of 5-dimensional anti-de~Sitter space and $SU(2,2|4)$ is the
superisometry of the $AdS_5\times S^5$ superspace as well as the $N=4$ extended superconformal symmetry of the $N=4$ supersymmetric Yang-Mills theory in the $D=4$ boundary superspace that provides the crucial
symmetry argument in support of the $AdS_5/CFT_4$ duality, it is
natural to seek for relevant (super)twistors. Because $AdS_5$
space-time is conformally flat, the incidence relations for
Penrose twistors (actually ambitwistors) admit natural extension
to accommodate extra bosonic coordinate(s) \cite{CRZ}, \cite{ASW}
(see also \cite{Cederwall}). It is also quite natural to add odd
coordinates for $D=4$ $N$-extended Poincare supersymmetry
\cite{ABGPKT16}, \cite{ABGPKT17}. The case of $AdS_5\times S^5$
superspace isomorphic to the $\coset$ supercoset manifold
\cite{MT'98}, \cite{Rahmfeld} is not so easy to deal with as it is
not superconformally flat \cite{BILS}. In \cite{Bars-ads5s5} there
was given the group-theoretical reasoning for the definition of
$AdS_5\times S^5$ supertwistors as two $SU(2)$ doublets of
$SU(2,2|4)$ supertwistors: one with even $SU(2,2)$ components and
odd $SU(4)$ components and another with odd $SU(2,2)$ components
and even $SU(4)$ ones. These constitute $(4|4)\times(2|2)$
rectangular block of the $\coset$ supermatrix.\footnote{Similar
ideas on the definition of supertwistors for $AdS_5/CFT_4$ duality
were also discussed in \cite{Siegel-twistor}.} We call such
supertwistors $c$- and $a$-type supertwistors by analogy with $c$-
and $a$-type supermatrices (see, e.g. \cite{Buchbinder}). Their
$SU(2,2|4)$-invariant products with dual supertwistors were shown
to satisfy seven bosonic and eight fermionic real constraints which number
matches that of $psu(2|2)\oplus u(1)$ generators.\footnote{Initial
version \cite{Bars-twistor-string} of the definition of
$AdS_5\times S^5$ supertwistors  used four $SU(2)$ doublets two of
which had even $SU(2,2)$ components and odd $SU(4)$ components and
other two had odd $SU(2,2)$ components and even $SU(4)$ components
and they satisfied the $s(u(2|2)\oplus
u(2|2))$ constraints. We believe that both definitions are
equivalent \cite{Uf}. The definition given in
\cite{Bars-twistor-string} relies on the spinor Lorentz harmonics
\cite{Galperin1}, \cite{Galperin2}, \cite{BZh} in $D=1+4$
dimensions parametrizing the coset $Spin(1,4)/(SU(2)\times SU(2))$,
that is to two $SU(2)$-Majorana spinors in $D=1+4$ dimensions, and
by solving corresponding harmonicity conditions one of them can be
expressed in terms of the other. Similarly the constraints on the
$D=5$ spinor harmonics parametrizing the coset
$Spin(5)/(SU(2)\times SU(2))$ can be solved in terms of one
$SU(2)$-Majorana spinor providing an evidence for the supertwistor
definition of \cite{Bars-ads5s5}.} However, the incidence
relations with the $AdS_5\times S^5$ supercoordinates, necessary,
for instance, to elaborate on the Penrose transform, have not been
obtained. Previous experience suggests that supertwistor
reformulation of the point-particle model is a proper tool to find
such relations. Additionally one of the advantages of the
supertwistor approach is that the superparticle's Lagrangian is
quadratic in supertwistors that facilitates quantization and can
give the supertwistor description of the $D=10$ $N=2$ chiral
supergravity spectrum compactified on  $AdS_5\times S^5$ \cite{GMarcus}, \cite{KRvN}.

Thus the aim of this note is to derive these $AdS_5\times S^5$
supertwistors starting from the first-order formulation of the
$D=1+9$ massless superparticle model in $AdS_5\times S^5$
superspace \cite{MTT}, \cite{Bars'02}, \cite{Horigane},
\cite{Siegel-superparticle}, \cite{Heinze}. We note that the
superparticle's momentum components tangent to $AdS_5$ and $S^5$
subspaces can be realized as the bilinears of the constrained
$SU(2)$-Majorana spinors in $D=1+4$ and $D=5$ dimensions
respectively. Contraction of their $SU(2)$ indices gives two
$4\times4$ traceless antisymmetric matrices that, when contracted
with the $(4|4)\times(4|4)$ supermatrix of the $psu(2,2|4)$ Cartan forms, extract
bosonic components of the supervielbein tangent to $AdS_5$ and
$S^5$ subspaces. This allows to bring the kinetic term in the
superparticle's Lagrangian to the form appropriate for the
introduction of the supertwistors. The details are worked out in
Section 2. Section 3 is devoted to the quantization of the
superparticle propagating in the $AdS_5$ subspace of the
$AdS_5\times S^5$ superspace both in the oscillator and
ambitwistor approaches that complement each other. In the simplest
case, when the particle's momentum tangent to $AdS_5$ is null, we
find how the $D=5$ $N=8$ gauged supergravity multiplet is embedded
into the ambitwistor superfield of homogeneity degree zero in each
of the arguments. Two appendices supply relevant details of the
spinor algebra and supermatrix realization of the generators of
$psu(2,2|4)$ isometry superalgebra of the $AdS_5\times S^5$
superspace.

\setcounter{equation}{0}
\section{Supertwistor mechanics of massless particle in $AdS_5\times S^5$ superspace}

\subsection{Superparticle moving on $AdS_5$ subspace: definition of $c$-type supertwistors}

Let us start with the group-theoretic consideration of the
$AdS_5\times S^5$ supervielbein components tangent to the $AdS_5$
space-time and $SO(1,4)$ connection 1-form. As is well known they
can be identified with the Cartan forms associated with the
generators of the $su(2,2)$ subalgebra of $psu(2,2|4)$. If $(4|4)\times(4|4)$
supermatrix $\mc G^{\mc A}{}_{\mc B}$ is a
$PSU(2,2|4)/(SO(1,4)\times SO(5))$ supercoset representative then
the left-invariant Cartan forms can be defined as
\beq\label{ads-cartan}
\mc G^{-1}d\mc G|_{\, su(2,2)}=\frac{i}{2}E_{\un m\un
n}(d)\tilde\rho^{\un m\un
n\bs\al}{}_{\bs\bt}=\frac{i}{2}E_{m'}(d)\g^{m'\bs\al}{}_{\bs\bt}+\frac{i}{2}E_{m'n'}(d)\g^{m'n'\bs\al}{}_{\bs\bt}\in
su(2,2),
\eeq
where restriction to the $su(2,2)$ subalgebra
amounts to focusing on the upper diagonal block of the $(4|4)\times(4|4)$
supermatrix of Cartan 1-forms, $E_{\un m\un n}(d)$ ($\un m,\un
n=0',0,1,2,3,5$) are the $su(2,2)$ Cartan 1-forms,
$E^{0'}{}_{m'}(d)=E_{m'}(d)$ ($m'=0,1,2,3,5$) are vielbein components
tangent to the anti-de Sitter space and $E_{m'n'}(d)$ is the
$SO(1,4)$ connection 1-form. $D=1+4$ $\g-$matrices
$\g^{m'\bs\al}{}_{\bs\bt}$ ($\bs\al,\bs\bt=1,2,3,4$) represent the
$su(2,2)$ generators from the $\mathfrak g_{(2)}$ eigenspace of
the $\mathbb Z_4$ outer automorphism of the $psu(2,2|4)$
superalgebra and their antisymmetrized products
$\g^{m'n'\bs\al}{}_{\bs\bt}$ represent generators from the $\mathfrak g_{(0)}$
eigenspace. In the considered realization important distinction
between these generators is that matrices
$\g^{m'}_{\bs\al\bs\bt}=C_{\bs\al\bs\g}\g^{m'\bs\g}{}_{\bs\bt}$
are antisymmetric, while
$\g^{m'n'}_{\bs\al\bs\bt}=C_{\bs\al\bs\g}\g^{m'n'\bs\g}{}_{\bs\bt}$
are symmetric in the spinor indices. Thus the product
$\la^i_{\bs\al}\la^{\bs\bt}_i$ ($i=1,2$) of two $D=1+4$
$SU(2)-$Majorana spinors
\beq
\la_i^{\bs\al}:\quad\overline{\la_i^{\bs\al}}=(\la^{\bs\bt}_i)^\dagger\g^{0\bs\bt}{}_{\bs\al}=(\la_{\bs\al}^i)^{\mathrm
T}
\eeq
acts as a projector onto the $\mathfrak g_{(2)}$
eigenspace. Promoting this $SU(2)$ doublet of $Spin(1,4)$ spinors
to 'superspinors'
\beq
\la^i_{\bs\al}\;\rightarrow\;\la^i_{\mc
A}=\left(
\begin{array}{c}
\la^i_{\bs\al} \\ 0
\end{array}
\right),\quad\la^{\bs\bt}_i\;\rightarrow\;\la^{\mc B}_i=\left(
\begin{array}{c}
\la_i^{\bs\bt} \\ 0
\end{array}
\right)
\eeq
one can write
\beq\label{ads-basic-relation}
\frac12\la^i_{\bs\al}\g^{m'\bs\al}{}_{\bs\bt}\la^{\bs\bt}_iE_{m'}(d)=-i\la^i_{\mc A}\mc G^{-1\,\mc A}{}_{\mc C}d\mc G^{\mc C}{}_{\mc B}\la^{\mc B}_i.
\eeq
This relation is the key to the twistor formulation of the superparticle on the $AdS_5$ subspace of the $AdS_5\times S^5$ superspace. 1-form on the r.h.s. of (\ref{ads-basic-relation}) can be presented in terms of $c-$type $AdS_5$ supertwistors
\beq
\la^i_{\mc A}\mc G^{-1\,\mc A}{}_{\mc C}d\mc G^{\mc C}{}_{\mc B}\la^{\mc B}_i=\frac12\left(\bar{\mc Z}^i_{\mc A}d\mc Z^{\mc A}_i-d\bar{\mc Z}^i_{\mc A}\mc Z^{\mc A}_i\right),
\eeq
where
\beq\label{c-supertwistor-definition}
\mc Z^{\mc A}_i=\mc G^{\mc A}{}_{\mc B}\la^{\mc B}_i,\quad\bar{\mc Z}^i_{\mc A}=\la^i_{\mc B}\mc G^{-1\,\mc B}{}_{\mc A}:\quad\bar{\mc Z}^i_{\mc A}=(\mc Z^{\mc B}_i)^\dagger\mc H^{\mc B}{}_{\mc A},
\eeq
and that on the l.h.s. of (\ref{ads-basic-relation}) defines the kinetic part of the superparticle's Lagrangian in the first-order form of the space-time formulation
\beq
\frac12\la^i_{\bs\al}\g_{m'}{}^{\bs\al}{}_{\bs\bt}\la^{\bs\bt}_iE^{m'}(d)=-p_{m'}E^{m'}(d)
\eeq
with
\beq\label{spinor-momentum}
p_{m'}=-\frac12\la^i_{\bs\al}\g_{m'}{}^{\bs\al}{}_{\bs\bt}\la^{\bs\bt}_i:\quad p^{m'}p_{m'}=-\frac14\Lambda^2,\quad\Lambda=\la^i_{\bs\al}\la^{\bs\al}_i.
\eeq
Vector $p^{m'}$ can be identified with the 5-momentum  of the massless or massive particle provided $\Lambda$ equals 0 or $m$.\footnote{$-m$ is also a possible option. For details see \cite{CKR}.} In the former case $\la^i_{\bs\al}$ can be identified with the $4\times2$ rectangular block of the $D=1+4$ spinor Lorentz-harmonic matrix that parametrizes the coset-space $Spin(1,4)/(SO(1,1)\times ISO(3))$ and $p^{m'}$ with the light-like vector-column from the respective vector Lorentz-harmonic matrix \cite{U'15}. For $m$ non-zero $\la^i_{\bs\al}$ can be related to the spinor Lorentz-harmonics parametrizing the coset $Spin(1,4)/SO(4)\sim Spin(1,4)/(SU(2)\times SU(2))$ and $p^{m'}$ with the vector-column of the respective vector Lorentz-harmonic matrix \cite{Uf}.
The net result is
\beq
p_{m'}E^{m'}(d)=\frac{i}{2}\left(\bar{\mc Z}^i_{\mc A}d\mc Z^{\mc A}_i-d\bar{\mc Z}^i_{\mc A}\mc Z^{\mc A}_i\right)
\eeq
for $p^{m'}$ satisfying (\ref{spinor-momentum}) and supertwistors constrained by the relations
\beq\label{c-supertwistor-constr}
\bar{\mc Z}^i_{\mc A}\mc Z^{\mc A}_j-\frac12\de^i_j\Lambda=0
\eeq
that provide the supertwistor realization of the generators of the $su(2)\oplus u(1)$ gauge algebra. So the Lagrangian of the superparticle moving in the $AdS_5$ subspace of the $AdS_5\times S^5$ superspace
\beq\label{ads-particle}
\mathscr L^{AdS_5}_{\mathrm{first-order}}=p_{m'}E^{m'}_\tau-\frac{g}{2}(p^{m'}p_{m'}+\Lambda^2)
\eeq
can be reformulated in terms of the $SU(2)$ doublet of $c-$type $SU(2,2|4)$ supertwistors
\beq\label{ads-twistor-particle}
\mathscr L^{AdS_5}_{\mathrm{supertwistor}}=\frac{i}{2}\left(\bar{\mc Z}^i_{\mc A}\dot{\mc Z}^{\mc A}_i-\dot{\bar{\mc Z}}\vp{\bar{\mc Z}}^i_{\mc A}\mc Z^{\mc A}_i\right)+a^{ij}\bar{\mc Z}_{\mc A i}\mc Z^{\mc A}_j+t(\bar{\mc Z}_{\mc A}^{i}\mc Z^{\mc A}_{i}-\Lambda),
\eeq
where $g$, $a^{ij}=a^{ji}$ and $t$ are the Lagrange multipliers. It can be checked that the number of the physical degrees of freedom in both formulations is the same.

To see how the incidence relations between the supertwistor
components and the coordinates of the $AdS_5\times S^5$ superspace
are encoded in (\ref{c-supertwistor-definition}) consider
definite $PSU(2,2|4)/(SO(1,4)\times SO(5))$ representative, e.g.
that discussed in \cite{MTlc}, \cite{MTT}, \cite{Metsaev'01}
\beq\label{coset-rep}
\mc G^{\mc A}{}_{\mc C}=\mc G_{AdS_5}{}^{\mc
A}{}_{\mc B}\mc G_{S^5}{}^{\mc B}{}_{\mc C} \eeq with \beq \mc
G_{AdS_5}=\exp(ix^mP_m+i\theta^\al_AQ^A_\al+i\bar\theta^{\dot\al
A}\bar Q_{\dot\al A})\exp(i\eta^B_\bt S^\bt_B+i\bar\eta_{\dot\bt
B}\bar S^{\dot\bt B})\exp(i\varphi D),
\eeq
corresponding to the
isomorphic realization of the $psu(2,2|4)$ superalgebra as the
$D=4$ $N=4$ superconformal algebra, and
\beq
\mc
G_{S^5}=\exp(iy^{I'}P^{I'}).
\eeq
Such choice of the supercoset
representative implies parametrization of
anti-de~Sitter space-time by the Poincare coordinates $x^m$
($m=0,1,2,3$), $\varphi$ with the line element \beq
ds^2_{AdS_5}=e^{-2\varphi}dx^mdx_m+d\varphi^2. \eeq $y^{I'}$
($I'=1,2,3,4,5$) are the $S^5$ coordinates, $\theta^\al_A$,
$\bar\theta^{\dot\al A}$ and $\eta^B_\bt$, $\bar\eta_{\dot\bt B}$
($\al,\bt=1,2$, $\dot\al,\dot\bt=1,2$, $A,B=1,2,3,4$) are odd
coordinates associated with the $N=4$ Poincare and conformal
supersymmetries respectively.

In the supermatrix realization of the relevant generators of $D=4$ $N=4$ superconformal algebra given in Appendix B
the first factor in (\ref{coset-rep}) acquires the form
\beq
\mc G_{AdS_5}{}^{\mc A}{}_{\mc B}=\left(
\begin{array}{ccc}
e^{\varphi/2}(\de^\al_\bt-2i\tilde x^{\dot\de\al}_+\bar\eta_{\dot\de D}\eta^D_\bt-4\theta^\al_D\eta^D_\bt) & ie^{-\varphi/2}\tilde x^{\dot\bt\al}_+ & 2i(\theta^\al_B+i\tilde x^{\dot\de\al}_+\bar\eta_{\dot\de B})\\[0.2cm]
-2e^{\varphi/2}\bar\eta_{\dot\al D}\eta^D_\bt & e^{-\varphi/2}\de^{\dot\bt}_{\dot\al} & 2i\bar\eta_{\dot\al B} \\[0.2cm]
2ie^{\varphi/2}(\eta^A_\bt-2\bar\theta^{\dot\de A}\bar\eta_{\dot\de D}\eta^D_\bt) & 2ie^{-\varphi/2}\bar\theta^{\dot\bt A} & \de^A_B-4\bar\theta^{\dot\de A}\bar\eta_{\dot\de B}
\end{array}
\right),
\eeq
where $\tilde x^{\dot\bt\al}_+=x^m\tilde\s^{\dot\bt\al}_m+2i\theta^\al_A\bar\theta^{\dot\bt A}$, and the second is
\beq
\mc G_{S^5}{}^{\mc B}{}_{\mc C}=\left(
\begin{array}{ccc}
\de^\bt_\g & 0 & 0\\[0.2cm]
0 & \de^{\dot\g}_{\dot\bt} & 0 \\[0.2cm]
0 & 0 & U^B_C
\end{array}
\right),\quad U^B_C=\cos|y|\de^B_C+i\frac{\sin|y|}{|y|}(\g\cdot y)^B{}_C.
\eeq

For such supercoset representative the $c-$type supertwistor incidence relations (\ref{c-supertwistor-definition}) acquire the form
\beq
\begin{array}{c}
\mc Z^{\mc A}_i=
\left(
\begin{array}{c}
\mu^{\al}_i \\[0.2cm]
\bar\Lambda_{\dot\al i} \\[0.2cm]
\eta^A_i
\end{array}
\right)=\mc G_{AdS_5}{}^{\mc A}{}_{\mc B}
\left(
\begin{array}{c}
-\la^\bt_i \\[0.2cm]
\bar\la_{\dot\bt i} \\[0.2cm]
0
\end{array}
\right)
\\[0.4cm]
=\left(
\begin{array}{c}
-e^{\varphi/2}\la^\al_i+ie^{-\varphi/2}\tilde x^{\dot\bt\al}_+\bar\la_{\dot\bt i}+4e^{\varphi/2}\theta^\al_B\eta^B_\bt\la^\bt_i+2ie^{\varphi/2}\tilde x^{\dot\bt\al}_+\bar\eta_{\dot\bt B}\eta^B_\bt\la^\bt_i \\[0.2cm]
e^{-\varphi/2}\bar\la_{\dot\al i}+2e^{\varphi/2}\bar\eta_{\dot\al B}\eta^B_\bt\la^\bt_i \\[0.2cm]
2ie^{-\varphi/2}\bar\theta^{\dot\bt A}\bar\la_{\dot\bt i}-2ie^{\varphi/2}\eta^A_\bt\la^\bt_i+4ie^{\varphi/2}\bar\theta^{\dot\bt A}\bar\eta_{\dot\bt B}\eta^B_\bt\la^\bt_i
\end{array}
\right).
\end{array}
\eeq
Accordingly for the dual supertwistor we have
\beq
\begin{array}{c}
\bar{\mc Z}^i_{\mc A}=(\Lambda^i_\al\;\bar\mu^{\dot\al i}\;\bar\eta^i_A)=(\la^i_\bt\;\bar\la^{\dot\bt i}\;0)\;\mc G^{-1}_{AdS_5}{}^{\mc B}{}_{\mc A}: \\[0.2cm]
\Lambda^i_\al=e^{-\varphi/2}\la^i_\al-2e^{\varphi/2}\bar\la^{\dot\bt i}\bar\eta_{\dot\bt D}\eta^D_\al,\\[0.2cm]
\bar\mu^{\dot\al i}=e^{\varphi/2}\bar\la^{\dot\al i}-ie^{-\varphi/2}\tilde x^{\dot\al\bt}_-\la^i_\bt-4e^{\varphi/2}\bar\la^{\dot\bt i}\bar\eta_{\dot\bt D}\bar\theta^{\dot\al D}+2ie^{\varphi/2}\bar\la^{\dot\bt i}\bar\eta_{\dot\bt D}\eta^D_\g\tilde x^{\dot\al\g}_-,\\[0.2cm]
\bar\eta^i_A=-2ie^{-\varphi/2}\la^i_\bt\theta^\bt_A-2ie^{\varphi/2}\bar\la^{\dot\bt i}\bar\eta_{\dot\bt A}+4ie^{\varphi/2}\bar\la^{\dot\bt i}\bar\eta_{\dot\bt D}\eta^D_\g\theta^\g_A,
\end{array}
\eeq
where
\beq
\mc G_{AdS_5}^{-1}{}^{\mc B}{}_{\mc A}=\left(
\begin{array}{ccc}
e^{-\varphi/2}\de^\bt_\al & -ie^{-\varphi/2}\tilde x^{\dot\al\bt}_- & -2ie^{-\varphi/2}\theta^\bt_A \\[0.2cm]
-2e^{\varphi/2}\bar\eta_{\dot\bt D}\eta^D_\al & e^{\varphi/2}(\de^{\dot\al}_{\dot\bt}+2i\bar\eta_{\dot\bt D}\eta^D_\g\tilde x^{\dot\al\g}_--4\bar\eta_{\dot\bt D}\bar\theta^{\dot\al D}) & -2ie^{\varphi/2}(\bar\eta_{\dot\bt A}-2\bar\eta_{\dot\bt D}\eta^D_\g\theta^\g_A) \\[0.2cm]
-2i\eta^B_\al & -2i(\bar\theta^{\dot\al B}-i\tilde x^{\dot\al\g}_-\eta_\g^B) & \de^B_A-4\eta^B_\g\theta^\g_A
\end{array}
\right)
\eeq
and the conjugation rules of the supertwistor components are $(\Lambda^i_\al)^\dagger=\bar\Lambda_{\dot\al i}$, $(\mu^\al_i)^\dagger=\bar\mu^{\dot\al i}$, $(\eta^A_i)^\dagger=\bar\eta^i_A$.

In the bosonic limit introduced supertwistors reduce to those proposed in \cite{CRZ} modulo the overall rescaling.  Penrose-Ferber supertwistors arise in the boundary limit $\varphi\to-\infty$
\beq
\mc Z^{\mc A}_i|_{\mathrm{boundary}}=\lim\limits_{\varphi\to-\infty}e^{\varphi/2}\mc Z^{\mc A}_i,\quad\bar{\mc Z}^i_{\mc A}|_{\mathrm{boundary}}=\lim\limits_{\varphi\to-\infty}e^{\varphi/2}\bar{\mc Z}^i_{\mc A}
\eeq
and we also call them boundary supertwistors.

\subsection{Definition of $a$-type supertwistors}

Now the above discussion can be generalized to the case of
superparticle moving also on the $S^5$ part of the
superbackground. Cartan forms associated with the generators of
the $su(4)$ subalgebra of $psu(2,2|4)$ decompose into the sum
\beq
\mc G^{-1}d\mc G|_{\,
su(4)}=\frac{i}{2}E^{IJ}(d)\tilde\rho^{IJA}{}_{B}=\frac{i}{2}E^{I'}(d)\g^{I'A}{}_{B}+\frac{i}{2}E^{I'J'}(d)\g^{I'J'A}{}_{B}\in
su(4),
\eeq
where restriction to the $su(4)$ subalgebra amounts to
considering the lower diagonal block of the $(4|4)\times(4|4)$ supermatrix of
Cartan forms, $E^{IJ}(d)$ ($I=1,2,3,4,5,6$) are the $su(4)$ Cartan
forms, 1-form $E^{6I'}(d)=E^{I'}(d)$ is identified with the $D=10$
supervielbein components tangent to $S^5$ and $E^{I'J'}(d)$ -- with
the $SO(5)$ connection. $D=5$ $\g-$matrices
$\g^{I'}_{AB}=C_{AD}\g^{I'D}{}_{B}$ are antisymmetric in the
spinor indices and belong to the $\mathfrak g_{(2)}$ eigenspace
under the $\mathbb Z_4$ automorphism of the $psu(2,2|4)$ superalgebra,
while the $so(5)$ generators realized by
$\g^{I'J'}_{AB}=C_{AD}\g^{I'J'D}{}_{B}$ are symmetric in the
spinor indices and belong to the $\mathfrak g_{(0)}$ eigenspace.
So that similarly to the $AdS_5$ case taking $D=5$
$SU(2)-$Majorana spinor $\ell^{i'}_A$ ($i'=1,2$):
\beq\label{d5-su2-majorana}
(\ell^{i'}_A)^\dagger=(\ell^A_{i'})^{\mathrm T}
\eeq
it is
possible to construct a matrix projector onto the $\mathfrak
g_{(2)}$ eigenspace. Presenting this spinor as a 'superspinor'
\beq \ell^{i'}_{\mc A}=\left(
\begin{array}{c}
0 \\[0.2cm] \ell^{i'}_A
\end{array}\right)
\eeq
yields
\beq\label{s5-projection}
\frac12\ell^{i'}_A\g^{I'A}{}_{B}\ell^B_{i'}E^{I'}(d)=-i\ell^{i'}_{\mc A}\mc G^{-1\,\mc A}{}_{\mc C}d\mc G^{\mc C}{}_{\mc B}\ell^{\mc B}_{i'}.
\eeq
Introducing $SU(2)$ doublet of $a-$type $SU(2,2|4)$ supertwistors
\beq\label{a-supertwistor}
\Psi^{\mc A}_{i'}=\mc G^{\mc A}{}_{\mc B}\ell^{\mc B}_{i'}
\eeq
and its dual
\beq\label{a-dual-supertwistor}
\bar\Psi^{i'}_{\mc A}=\ell^{i'}_{\mc B}\mc G^{-1\,\mc B}{}_{\mc A}:\quad\bar\Psi^{i'}_{\mc A}=(\Psi^{\mc B}_{i'})^\dagger\mc H^{\mc B}{}_{\mc A},
\eeq
projected Cartan 1-form on the r.h.s. of (\ref{s5-projection}) can be written as
\beq
\ell^{i'}_{\mc A}\mc G^{-1\,\mc A}{}_{\mc C}d\mc G^{\mc C}{}_{\mc B}\ell^{\mc B}_{i'}=\frac12(\bar\Psi^{i'}_{\mc A}d\Psi^{\mc A}_{i'}-d\bar\Psi^{i'}_{\mc A}\Psi^{\mc A}_{i'}).
\eeq
Further introducing 5-vector
\beq\label{sphere-momentum}
p^{I'}=-\frac12\ell^{i'}_A\g^{I'A}{}_{B}\ell^B_{i'},
\eeq
(\ref{s5-projection}) acquires the form
\beq
p^{I'}E^{I'}(d)=\frac{i}{2}(\bar\Psi^{i'}_{\mc A}d\Psi^{\mc A}_{i'}-d\bar\Psi^{i'}_{\mc A}\Psi^{\mc A}_{i'}).
\eeq
The norm of this vector equals
\beq
p^{I'}p^{I'}=L^2,\quad L=\frac14\ell^{i'}_A\ell^A_{i'}
\eeq
and supertwistors (\ref{a-supertwistor}), (\ref{a-dual-supertwistor}) satisfy four constraints
\beq\label{a-supertwistor-constr}
\bar\Psi^{i'}_{\mc A}\Psi^{\mc A}_{j'}-\frac12\de^{i'}_{j'}L=0.
\eeq

Like for the $c$-type supertwistors we can give explicit form of the incidence relations for $a-$type supertwistors (\ref{a-supertwistor}) and their duals (\ref{a-dual-supertwistor}) for the $PSU(2,2|4)/(SO(1,4)\times SO(5))$ supercoset representative (\ref{coset-rep})
\beq
\Psi^{\mc A}_{i'}=\left(
\begin{array}{c}
m^\al_{i'} \\[0.2cm]
\bar\chi_{\dot\al i'} \\[0.2cm]
L^A_{i'}
\end{array}
\right)=\mc G^{\mc A}{}_{\mc B}\left(
\begin{array}{c}
0 \\[0.2cm]
0 \\[0.2cm]
\ell^B_{i'}
\end{array}
\right)=\left(
\begin{array}{c}
2i(\theta^\al_B+i\tilde x^{\dot\bt\al}_+\bar\eta_{\dot\bt B})\tilde\ell^B_{i'} \\[0.2cm]
2i\bar\eta_{\dot\al B}\tilde\ell^B_{i'} \\[0.2cm]
(\de^A_B-4\bar\theta^{\dot\bt A}\bar\eta_{\dot\bt B})\tilde\ell^B_{i'}
\end{array}
\right)
\eeq
and
\beq
\begin{array}{c}
\bar\Psi^{i'}_{\mc A}=(\chi_{\al}^{i'}\;\bar m^{\dot\al i'}\; \bar L^{i'}_A)=(0\;0\;\ell^{i'}_B)\mc G^{-1}{}^{\mc B}{}_{\mc A}:\\[0.2cm]
\chi_{\al}^{i'}=-2i\bar{\tilde\ell}^{i'}_B\eta^{B}_\al,\quad
%\\[0.2cm]
\bar m^{\dot\al i'}=-2i\bar{\tilde\ell}^{i'}_B(\bar\theta^{\dot\al B}-i\tilde x^{\dot\al\bt}_-\eta^{B}_\bt),\quad
%\\[0.2cm]
\bar L^{i'}_A=\bar{\tilde\ell}^{i'}_B(\de^B_A-4\eta^B_\bt\theta^\bt_A),
\end{array}
\eeq
where $\tilde\ell^B_{i'}=U^B{}_C\ell^C_{i'}$, $\bar{\tilde\ell}^{i'}_B=\ell^{i'}_C\bar U^C{}_B$ and the component form of the conjugation rules in (\ref{a-dual-supertwistor}) is $(m^\al_{i'})^\dagger=\bar m^{\dot\al i'}$, $(\chi_{\al}^{i'})^\dagger=\bar\chi_{\dot\al i'}$, $(L^A_{i'})^\dagger=\bar L^{i'}_A$. Observe that components of $\Psi^{\mc A}_{i'}$ and $\bar\Psi^{i'}_{\mc A}$ supertwistors depend not only on the $S^5$ coordinates $y^{I'}$ but also on $\tilde x^{\dot\al\bt}$ so that in general it is not possible to view $c-$ and $a-$type supertwistors as related solely to the $AdS_5$ and $S^5$ bosonic subspaces of the $AdS_5\times S^5$ superspace. Also note that $c-$ and $a-$type supertwistors are by definition mutually orthogonal
\beq\label{a-c-orthogonal}
\bar\Psi^{i'}_{\mc A}\mc Z^{\mc A}_i=0,\quad\bar{\mc Z}^i_{\mc A}\Psi^{\mc A}_{i'}=0.
\eeq

\subsection{Classical formulation of massless particle in $AdS_5\times S^5$ superspace in terms of $c$- and $a$-type supertwistors}

The above results can be put together to describe the $D=1+9$ massless superparticle on the $AdS_5\times S^5$ superbackground. First, its null 10-momentum $p^{\hat m}$ can be assembled from 5-vectors (\ref{spinor-momentum}) and (\ref{sphere-momentum})
\beq\label{10-momentum}
p^{\hat m}=(p^{m'},\, p^{I'}):\quad p^{\hat m}p_{\hat m}=0\,\rightarrow\,\Lambda=\pm L.
\eeq
The sign ambiguity is resolved by requiring the closure of the algebra of the first-class constraints. Then the first-order Lagrangian (\ref{ads-particle})  of the superparticle propagating on the $AdS_5$ subspace of $AdS_5\times S^5$ space generalizes to
\beq
\mathscr L^{AdS_5\times S^5}_{\mathrm{first-order}}=p_{m'}E_\tau^{m'}+p^{I'}E^{I'}_\tau-\frac{g}{2}(p^{m'}p_{m'}+p^{I'}p^{I'})
\eeq
and its supertwistor version acquires the form
\beq
\begin{array}{rl}
\mathscr L^{AdS_5\times S^5}_{\mathrm{supertwistor}}=&\frac{i}{2}\left(\bar{\mc Z}^i_{\mc A}\dot{\mc Z}^{\mc A}_i-\dot{\bar{\mc Z}}\vp{\bar{\mc Z}}^i_{\mc A}\mc Z^{\mc A}_i\right)+\frac{i}{2}(\bar\Psi^{i'}_{\mc A}\dot\Psi^{\mc A}_{i'}-\dot{\bar\Psi}^{i'}_{\mc A}\Psi^{\mc A}_{i'}) \\[0.2cm]
+&a^{ij}\bar{\mc Z}_{\mc A i}\mc Z^{\mc A}_j+s^{i'j'}\bar\Psi_{\mc A i'}\Psi^{\mc A}_{j'}+t(\bar{\mc Z}_{\mc A}^{i}\mc Z^{\mc A}_{i}\pm\bar\Psi_{\mc A}^{i'}\Psi^{\mc A}_{i'}) \\[0.2cm]
+&i\bar\La^i_{i'}\bar\Psi^{i'}_{\mc A}\mc Z^{\mc A}_i+i\La^{i'}_i\bar{\mc Z}^i_{\mc A}\Psi^{\mc A}_{i'}.
\end{array}
\eeq
Lagrange multipliers $a^{ij}=a^{ji}$, $s^{i'j'}=s^{j'i'}$ and $t$ are even and $\varkappa^i_{i'}$, $\bar\varkappa^{i'}_i$ are odd. They introduce the first-class constraints
\beq\label{constraints-full}
\begin{array}{c}
L_i{}^j=\bar{\mc Z}^j_{\mc A}\mc Z^{\mc A}_i-\frac12\de^j_i\bar{\mc Z}^k_{\mc A}\mc Z^{\mc A}_k\approx0, \\[0.2cm]
M_{i'}{}^{j'}=\bar\Psi^{j'}_{\mc A}\Psi^{\mc A}_{i'}-\frac12\de^{j'}_{i'}\bar\Psi^{k'}_{\mc A}\Psi^{\mc A}_{k'}\approx0, \\[0.2cm]
T_\pm=\bar{\mc Z}_{\mc A}^{i}\mc Z^{\mc A}_{i}\pm\bar\Psi_{\mc A}^{i'}\Psi^{\mc A}_{i'}\approx0, \\[0.2cm]
\Phi^{i'}_i=\bar\Psi^{i'}_{\mc A}\mc Z^{\mc A}_i\approx0,\quad\bar\Phi^{i}_{i'}=\bar{\mc Z}^i_{\mc A}\Psi^{\mc A}_{i'}\approx0.
\end{array}
\eeq
On the Dirac brackets (D.B.)
\begin{eqnarray}
\{\mc Z^{\mc A}_i,\bar{\mc Z}^j_{\mc B}\}_{D.B.}&=&i\de^{\mc A}_{\mc B}\de_i^j, \label{db-c} \\[0.2cm]
\{\Psi^{\mc A}_{i'},\bar\Psi^{j'}_{\mc B}\}_{D.B.}&=&i\de^{\mc A}_{\mc B}\de_{i'}^{j'} \label{db-a}
\end{eqnarray}
constraints (\ref{constraints-full}) satisfy the $psu(2|2)\oplus u(1)$ superalgebra relations
\beq
\begin{array}{c}
\{L_i{}^j,L_k{}^l\}_{D.B.}=i\de_i^lL_k{}^j-i\de^j_kL_i{}^l,\quad\{M_{i'}{}^{j'},M_{k'}{}^{l'}\}_{D.B.}=i\de_{i'}^{l'}M_{k'}{}^{j'}-i\de_{k'}^{j'}M_{i'}{}^{l'}; \\[0.2cm]
\{\Phi^{i'}_i,\bar\Phi^{j}_{j'}\}_{D.B.}=i\de^{i'}_{j'}L_i{}^j+i\de^j_iM_{j'}{}^{i'}+\frac{i}{2}\de_i^j\de^{i'}_{j'}T_+; \\[0.2cm]
\{L_i{}^j,\Phi^{k'}_k\}_{D.B.}=-i\de^j_k\Phi^{k'}_i+\frac{i}{2}\de^i_j\Phi^{k'}_k,\quad
\{L_i{}^j,\bar\Phi^{k}_{k'}\}_{D.B.}=i\de^k_i\bar\Phi^{j}_{k'}-\frac{i}{2}\de^j_i\bar\Phi^{k}_{k'}, \\[0.2cm]
\{M_{i'}{}^{j'},\Phi^{k'}_k\}_{D.B.}=i\de^{k'}_{i'}\Phi^{j'}_k-\frac{i}{2}\de^{j'}_{i'}\Phi^{k'}_k,\quad
\{M_{i'}{}^{j'},\bar\Phi^{k}_{k'}\}_{D.B.}=-i\de^{j'}_{k'}\bar\Phi^{k}_{i'}+\frac{i}{2}\de^{j'}_{i'}\bar\Phi^{k}_{k'}.
\end{array}
\eeq
Above D.B. relations show that $L_i{}^j$ and $M_{i'}{}^{j'}$ generate two copies of the $su(2)$ algebra and $\Phi^{i'}_i$, $\bar\Phi^{i}_{i'}$ are the supersymmetry generators. Altogether they span $psu(2|2)$ superalgebra. The presence of $T_+$ on the r.h.s. of the anticommutation relations of the supersymmetry generators necessitates its inclusion into the set of the first-class constraints and fixes the sign in (\ref{10-momentum}) and (\ref{constraints-full}). The real dimension of the first-class constraint set $(7|8)$ is such that the number of the physical degrees of freedom of the superparticle matches both in the supertwistor and superspace formulations.

Let us finally discuss the gauge symmetries of the $D=1+9$ massless superparticle action in the supertwistor formulation
\beq\label{action}
S=\int d\tau\mathscr L^{AdS_5\times S^5}_{\mathrm{supertwistor}}
\eeq
that are generated by the first-class constraints (\ref{constraints-full}). $L_i{}^j$ and $M_{i'}{}^{j'}$ are the generators of the local $SU(2)$ rotations of each doublet of the supertwistors
\beq
\begin{array}{c}
\de_{\mathrm a}{\mc Z}^{\mc A}_i=-i\mathrm a_i{}^j{\mc Z}^{\mc A}_j,\quad\de_{\mathrm a}\bar{\mc Z}^i_{\mc A}=i\bar{\mc Z}^j_{\mc A}\mathrm a_j{}^i,\\[0.2cm]
\de_{\mathrm s}\Psi^{\mc A}_{i'}=-i\mathrm s_{i'}{}^{j'}\Psi^{\mc A}_{j'},\quad\de_{\mathrm s}\bar\Psi^{i'}_{\mc A}=i\bar\Psi^{j'}_{\mc A}\mathrm s_{j'}{}^{i'}.
\end{array}
\eeq
with the parameters $\mathrm a_i{}^j(\tau)$ and $\mathrm s_{i'}{}^{j'}(\tau)$. Invariance of the action (\ref{action}) requires the Lagrange multipliers to have non-zero variations
\beq
\begin{array}{c}
\de_{\mathrm a}\La^j{}_i=-\dot{\mathrm a}_i{}^j+i(\La^k{}_i\mr a_k{}^j-\La^j{}_k\mr a_i{}^k),\quad\de_{\mathrm a}\La^{i'}_i=-i\mr a_i{}^j\La_j^{i'},\quad\de_{\mathrm a}\bar\La_{i'}^i=i\bar\La_{i'}^j\mr a_j{}^i, \\[0.2cm]
\de_{\mathrm s}\La^{j'}{}_{i'}=-\dot{\mathrm s}_{i'}{}^{j'}+i(\La^{k'}{}_{i'}s_{k'}{}^{j'}-\La^{j'}{}_{k'}s_{i'}{}^{k'}),\quad\de_{\mathrm s}\La^{i'}_i=i\La^{j'}_is_{j'}{}^{i'},\quad\de_{\mathrm s}\bar\La^i_{i'}=-is_{i'}{}^{j'}\bar\La^i_{j'}.
\end{array}
\eeq
The constraint
$T'\approx0$
generates phase rotation of the supertwistors
\beq
\de_{\mathrm t}\mc Z^{\mc A}_i=i\mathrm t\mc Z^{\mc A}_i,\quad\de_{\mathrm t}\bar{\mc Z}^i_{\mc A}=-i\mathrm t\bar{\mc Z}^i_{\mc A},\quad
\de_{\mathrm t}\Psi^{\mc A}_{i'}=i\mathrm t\Psi^{\mc A}_{i'},\quad\de_{\mathrm t}\bar\Psi^{i'}_{\mc A}=-i\mathrm t\bar\Psi^{i'}_{\mc A},
\eeq
where $\mathrm t(\tau)$ is the local parameter. It has to be accompanied by the variation $\de_{\mathrm t}\La=\dot{\mathrm t}$ for the action invariance. Odd constraints
%$\Phi^{i'}_i$ and $\bar\Phi^{i}_{i'}$
generate the 8-parametric remnant of the $\kappa-$symmetry of the superparticle action in the superspace formulation with the parameters $\kappa^{i'}_i(\tau)$ and $\bar\kappa^i_{i'}(\tau)$
\beq\label{kappa}
\begin{array}{c}
\de_{\kappa}\mc Z^{\mc A}_i=(-)^{\varepsilon(\Psi)}\kappa^{i'}_i\Psi^{\mc A}_{i'},\quad\de_{\kappa}\bar{\mc Z}^i_{\mc A}=\bar\kappa^i_{i'}\bar\Psi^{i'}_{\mc A},\quad\de_\kappa\bar\Psi^{i'}_{\mc A}=\kappa^{i'}_i\bar{\mc Z}^i_{\mc A},\quad\de_{\kappa}\Psi^{\mc A}_{i'}=(-)^{\varepsilon(\mc Z)}\bar\kappa^i_{i'}\mc Z^{\mc A}_i, \\[0.2cm]
\de_\kappa\La^{i'}_i=\dot\kappa^{i'}_i-i(\kappa^{i'}_j\La^j{}_i-\La^{i'}{}_{j'}\kappa^{j'}_i),\quad\de_{\kappa}\bar\La^i_{i'}=\dot{\bar\kappa}^i_{i'}+i(\La^i{}_j\bar\kappa^j_{i'}-\bar\kappa^i_{j'}\,\La^{j'}{}_{i'}),\\[0.2cm]
\de_\kappa\La^i{}_j=i(\bar\kappa^i_{i'}\La^{i'}_j-\bar\La^i_{i'}\kappa^{i'}_j),\quad\de_\kappa\La^{j'}{}_{i'}=i(\kappa^{j'}_i\bar\La^i_{i'}-\La^{j'}_i\bar\kappa^i_{i'}),\quad\de_\kappa\La=\frac{i}{2}(\bar\kappa^i_{i'}\La^{i'}_i+\kappa^{i'}_i\bar\La^i_{i'}).
\end{array}
\eeq
$\varepsilon(\mc Z)$ and $\varepsilon(\Psi)$ equal 0 for bosonic components and 1 for fermionic components of respective supertwistors.

\setcounter{equation}{0}
\section{Aspects of quantum theory of massless superparticle in $AdS_5$ subspace of $AdS_5\times S^5$ superspace}

In this section we consider quantum theory of the superparticle moving in the $AdS_5$ subspace of $AdS_5\times S^5$ superspace. Quantization is performed both in terms of the oscillators and supertwistors that yields two (complementary) views on the $D=5$ $N=8$ gauged supergravity multiplet.

\subsection{Oscillator quantization}

There is well-known intimate relation between the $SU(2)$
oscillators and ($c$-type) supertwistors \cite{GMZ1},
\cite{CGKRZ}. For the bosonic components of supertwistors it is
based on two forms of the $SU(2,2)$ 'metric' that connects
fundamental and antifundamental representations. It is
off-diagonal in the (super)twistor basis and diagonal in the
oscillator basis. Thus the $SU(2)$ bosonic oscillators are defined
by the linear combinations of the supertwistor bosonic components
\beq
a^{\al}=\frac{1}{\sqrt{2}}(-\mu^\al+\bar\Lambda^\al),\quad
a_\al=\frac{1}{\sqrt{2}}(-\bar\mu_\al+\Lambda_\al)
\eeq
and
\beq
b_{\al}=\frac{1}{\sqrt{2}}(\bar\mu_\al+\Lambda_\al),\quad
b^\al=\frac{1}{\sqrt{2}}(\mu^\al+\bar\Lambda^\al).
\eeq
The
indices of internal $SU(2)$ symmetry of the supertwistors have
been suppressed for the moment. In quantum theory these
oscillators satisfy the commutation relations
\beq\label{oscillators}
[a_\al,a^\bt]=\de_\al^\bt,\quad
[b^\al,b_\bt]=\de^\al_\bt
\eeq
which can be deduced from the
quantum counterpart of the D.B. relations in (\ref{db-c}). Relations (\ref{oscillators}) suggest interpretation of $a^\al$ and $b_\al$ as raising oscillators and of $a_\al$ and $b^\al$ as lowering oscillators.

Since the definition of $SU(2,2|4)$ supertwistors assumes that $SU(4)$ 'metric' is unit, fermionic $SU(2)$ oscillators can be identified with odd components of supertwistors
\beq
\eta^A=\left(
\begin{array}{c}
\al^a \\ \bt^{\dot a}
\end{array}
\right),\qquad
\bar\eta_A=\left(
\begin{array}{c}
\al_a \\ \bt_{\dot a}
\end{array}
\right),
\eeq
where we used the decomposition of the $SU(4)$ (anti)fundamental representation $\mathbf4=\mathbf2\oplus\tilde{\mathbf2}$ ($\bar{\mathbf4}=\bar{\mathbf2}\oplus\tilde{\bar{\mathbf2}}$). Fermionic oscillators satisfy anticommutation relations
\beq
\{\al^a,\al_b\}=\de^a_b,\quad\{\bt^{\dot a},\bt_{\dot b}\}=\de^{\dot a}_{\dot b}
\eeq
as follows from (\ref{db-c}). It is conventional to treat $\al^a$ and $\bt_{\dot a}$ as raising oscillators, while $\al_a$ and $\bt^{\dot a}$ as lowering ones.

Now with each of the conjugate pairs of supertwistors (\ref{c-supertwistor-definition}) associate a set of the above introduced bosonic and fermionic oscillators
\beq
\begin{array}{c}
(\mc Z^{\mc A}_1,\bar{\mc Z}^1_{\mc A})\,\to\,(a^\al(1),b^\al(1),\al^a(1),\bt^{\dot a}(1);a_\al(1),b_\al(1),\al_{a}(1),\bt_{\dot a}(1)), \\[0.2cm]
(\mc Z^{\mc A}_2,\bar{\mc Z}^2_{\mc A})\,\to\,(a^\al(2),b^\al(2),\al^a(2),\bt^{\dot a}(2);a_\al(2),b_\al(2),\al_{a}(2),\bt_{\dot a}(2)).
\end{array}
\eeq
Then the $su(2)\oplus u(1)$ constraints (\ref{c-supertwistor-constr}) can be brought to the following form in terms of the oscillators
\beq\label{osc-constraints}
\begin{array}{c}
-N_{(a)}(1)+N_{(b)}(1)-N_{(\al)}(1)+N_{(\bt)}(1)-\frac12\Lambda\approx0,\\[0.2cm]
-N_{(a)}(2)+N_{(b)}(2)-N_{(\al)}(2)+N_{(\bt)}(2)-\frac12\Lambda\approx0,\\[0.2cm]
-a_\al(2)a^\al(1)+b_\al(2)b^\al(1)+\al_a(2)\al^a(1)+\bt_{\dot a}(2)\bt^{\dot a}(1)\approx0,\\[0.2cm]
-a_\al(1)a^\al(2)+b_\al(1)b^\al(2)+\al_a(1)\al^a(2)+\bt_{\dot a}(1)\bt^{\dot a}(2)\approx0,
\end{array}
\eeq
where $N_{(a)}(1)=a^\al(1)a_\al(1)$,
$N_{(b)}(1)=b_\al(1)b^\al(1)$, $N_{(\al)}(1)=\al^a(1)\al_a(1)$ and
$N_{(\bt)}(1)=\bt_{\dot a}(1)\bt^{\dot a}(1)$ are the oscillator
number operators for bosonic and fermionic oscillators of the
first set. Analogous definitions apply to the second set of
oscillators. Important feature of the constraints
(\ref{osc-constraints}) is that they commute with the raising
supersymmetry operators
\beq
a^\al(1)\bt_{\dot
a}(1)+a^\al(2)\bt_{\dot a}(2),\quad
b_\al(1)\al^a(1)+b_\al(2)\al^a(2)
\eeq
used to generate
lowest-weight vectors corresponding to different $SU(2,2)\times
SU(4)$ representations within the same $SU(2,2|4)$ supermultiplet
as well as with the bosonic
\beq
a^\bt(1)b_\al(1)+a^\bt(2)b_\al(2)
\eeq
and fermionic
\beq
\al^b(1)\bt_{\dot a}(1)+\al^b(2)\bt_{\dot
a}(2)
\eeq
raising operators which applied to a lowest-weight
vector generate the whole set of basis vectors of the
corresponding $SU(2,2)\times SU(4)$ representation \cite{GMarcus},
\cite{GMZ2}. So  one can consider the constraints
(\ref{osc-constraints}) to act on the $SU(2,2)\times SU(4)$ lowest-weight
vectors. In the simplest case $\Lambda=0$, i.e. when the
superparticle does not have non-zero momentum components in the directions tangent to
$S^5$, it is not hard to verify that the only
lowest-weight vector annihilated by the constraints is the
oscillator vacuum associated with the $D=5$ $N=8$ gauged
supergravity multiplet \cite{GMarcus}. For $\Lambda\not=0$ the
constraints select $SU(2)\times U(1)$ invariant lowest-weight
vectors which correspond to the supermultiplets discussed in
\cite{GMZ1} (see also \cite{CKR}).

\subsection{Ambitwistor quantization}

Quantum counterpart of the D.B. relations (\ref{db-c})
\beq
[\mc Z^{\mc A}_i,\bar{\mc Z}^j_{\mc B}\}=\de^{\mc A}_{\mc B}\de_i^j
\eeq
is consistent with the well-known in twistor theory \cite{MCP}, \cite{Shirafuji} realizations of quantized (super)twistors as multiplication and differentiation operators. For the ambitwistor space description of the quantized superparticle the supertwistors
\beq
\mc Z^{\mc A}_1=\left(
\begin{array}{c}
Z^{\boldsymbol{\al}}_1 \\ \eta^A_1
\end{array}
\right)
\equiv
\mc Z^{\mc A}=\left(
\begin{array}{c}
Z^{\boldsymbol{\al}} \\ \eta^A
\end{array}
\right),
\quad
\bar{\mc Z}^1_{\mc A}=(\bar Z_{\boldsymbol{\al}}^1,\:\bar\eta^1_A)\equiv
\bar{\mc Z}_{\mc A}=(\bar Z_{\boldsymbol{\al}},\:\bar\eta_A)
\eeq
and
\beq
\mc Z^{\mc A}_2=\left(
\begin{array}{c}
Z^{\boldsymbol{\al}}_2 \\ \eta^A_2
\end{array}
\right)
\equiv
\mc W^{\mc A}=
\left(
\begin{array}{c}
W^{\boldsymbol{\al}} \\ \zeta^A
\end{array}
\right),\quad \bar{\mc Z}^2_{\mc A}=(\bar
Z_{\boldsymbol{\al}}^2,\:\bar\eta^2_A)\equiv\bar{\mc W}_{\mc A}
=(\bar W_{\boldsymbol{\al}},\:\bar\zeta_A)
\eeq
can be conveniently
realized (modulo their relabeling) as
\beq\label{qtwistor-realization}
\begin{array}{c}
Z^{\boldsymbol{\al}}\rightarrow Z^{\boldsymbol{\al}},\quad\bar Z_{\boldsymbol{\al}}\rightarrow-\frac{\pt}{\pt Z^{\boldsymbol{\al}}},\quad\eta^A\rightarrow\eta^A,\quad\bar\eta_A\rightarrow\frac{\vec\pt}{\pt\vp{\hat\eta^A}\eta^A} \\[0.2cm]
W^{\boldsymbol{\al}}\rightarrow\frac{\pt}{\pt\vp{\hat W_{\boldsymbol{\al}}}\bar W_{\boldsymbol{\al}}},\quad\bar W_{\boldsymbol{\al}}\rightarrow\bar W_{\boldsymbol{\al}},\quad\zeta^A\rightarrow\frac{\vec\pt}{\pt\vp{\breve\zeta_A}\bar\zeta_A},\quad\bar\zeta_A\rightarrow\bar\zeta_A,
\end{array}
\eeq
where odd derivatives are defined to act from the left. Then in
the simplest case $\Lambda=0$ the $su(2)\oplus u(1)$ constraints
(\ref{c-supertwistor-constr}) applied to the superparticle's wave
function acquire the form \beq\label{fix-hom-degree}
\begin{array}{c}
\mc Z^{\mc A}\frac{\pt}{\pt\vp{\mc Z^{\hat{\mc A}}}\mc Z^{\mc A}}F_{(0,0)}(\mc Z,\bar{\mc W})=\left(Z^{\bs{\al}}\,\frac{\pt}{\pt\vp{Z^{\hat A}} Z^{\bs{\al}}}+\eta^{A}\frac{\pt}{\pt\vp{\eta^{\hat A}}\eta^A}\right)F_{(0,0)}(\mc Z,\bar{\mc W})=0,\\[0.3cm]
\bar{\mc W}_{\mc A}\frac{\pt}{\pt\vp{\hat{\mc W}}\bar{\mc W}_{\mc A}}F_{(0,0)}(\mc Z,\bar{\mc W})=\left(\bar W_{\bs{\al}}\,\frac{\pt}{\vp{\hat W}\pt\bar W_{\bs{\al}}}+\bar\zeta_A\frac{\pt}{\vp{\hat\zeta}\pt\bar\zeta_A}\right)F_{(0,0)}(\mc Z,\bar{\mc W})=0
\end{array}
\eeq
and
\beq\label{mult-delta-function}
\bar{\mc W}_{\mc A}\mc Z^{\mc A}F_{(0,0)}(\mc Z,\bar{\mc W})=0.
\eeq
Eqs.~(\ref{fix-hom-degree}) imply that $F_{(0,0)}$ has homogeneity degree zero both in $\mc Z$ and $\bar{\mc W}$ that is indicated by its subscripts and the condition (\ref{mult-delta-function}) can be taken into account by adding a $\de$-function factor
%representing $F_{(0,0)}$ in the following form
\beq\label{delta-function}
F_{(0,0)}(\mc Z,\bar{\mc W})=\de(\bar{\mc W}\mc Z)f_{(1,1)}(\mc Z,\bar{\mc W}).
\eeq
The function $f_{(1,1)}(\mc Z,\bar{\mc W})$ has the power series decomposition in odd supertwistor components
\beq\label{wf-decomposition}
\begin{array}{rl}
f_{(1,1)}(\mc Z,\bar{\mc W})=&h_{(1,1)}(Z,\bar W)+\psi_{(0,1)A}(Z,\bar W)\eta^A+\psi_{(1,0)}{}^A(Z,\bar W)\bar\zeta_A\\[0.2cm]
+&b_{(-1,1)}{}_{[AB]}(Z,\bar W)\eta^A\eta^B+a_{(0,0)}{}^B_A(Z,\bar W)\eta^A\bar\zeta_B+b_{(1,-1)}{}^{[AB]}(Z,\bar W)\bar\zeta_A\bar\zeta_B\\[0.2cm]
+&\lambda_{(-2,1)[ABC]}(Z,\bar W)\eta^A\eta^B\eta^C+\lambda_{(-1,0)}{}^C_{[AB]}(Z,\bar W)\eta^A\eta^B\bar\zeta_C\\[0.2cm]
+&\lambda_{(0,-1)}{}^{[BC]}_{A}(Z,\bar W)\eta^A\bar\zeta_B\bar\zeta_C+\lambda_{(1,-2)}{}^{[ABC]}(Z,\bar W)\bar\zeta_A\bar\zeta_B\bar\zeta_C\\[0.2cm]
+&\varphi_{(-3,1)[ABCD]}(Z,\bar W)\eta^A\eta^B\eta^C\eta^D+\varphi_{(-2,0)}{}_{[ABC]}^D(Z,\bar W)\eta^A\eta^B\eta^C\bar\zeta_D\\[0.2cm]
+&\varphi_{(-1,-1)}{}_{[AB]}^{[CD]}(Z,\bar W)\eta^A\eta^B\bar\zeta_C\bar\zeta_D+\varphi_{(0,-2)}{}_A^{[BCD]}(Z,\bar W)\eta^A\bar\zeta_B\bar\zeta_C\bar\zeta_D+\\[0.2cm]
+&\varphi_{(1,-3)}{}^{[ABCD]}(Z,\bar W)\bar\zeta_A\bar\zeta_B\bar\zeta_C\bar\zeta_D+\ldots
\end{array}
\eeq
where dots stand for higher terms in the series that as we shall see below are irrelevant.

Since the Penrose transform between the homogeneous functions of
the introduced twistors and the fields on $AdS_5\times S^5$
superspace is yet to be elaborated,\footnote{See, however,
Ref.~\cite{ASW}, where, based on the realization of $AdS_5$ space
as a projective manifold, the Penrose transform for the case of
spin 0 and $1/2$ fields was considered.} we can meanwhile consider
the Penrose transform of the corresponding functions of the
boundary ambitwistors that should produce the fields of $D=4$
$N=4$ conformal supergravity multiplet
\cite{FT'85}.\footnote{Hereinafter the same notation introduced
above for the bulk ambitwistors is used for the boundary ones. In
the Poincare coordinates taking the boundary limit of $AdS_5$
supertwistors is accompanied by their rescaling but since the
superparticle's wave function is scale-invariant in both arguments
the relations (\ref{delta-function}) and (\ref{wf-decomposition})
are valid as they stand either for bulk or boundary
supertwistors.} According to the AdS/CFT dictionary these fields
of the $D=4$ $N=4$ conformal supergravity multiplet correspond to
the boundary values of non-normalizable solutions of the bulk
$D=5$ $N=8$ gauged supergravity equations linearized around
$AdS_5$ \cite{Liu}, \cite{Balasubramanian}.\footnote{General
discussion of the correspondence between the $AdS$ bulk gauge fields and
the boundary conformal (shadow) fields not restricted to the case
of low spins may be found in \cite{Metsaev08}, \cite{Metsaev13}.}
Thus the function $h_{(1,1)}(Z,\bar W)$
%, $a_{(0,0)}{}^B_A(Z,\bar W)$ and $\varphi_{(-1,-1)}{}_{[AB]}^{[CD]}(Z,\bar W)$
yields Minkowski space graviton field $h_{\al(2)\dot\al(2)}(x)$ (see, e.g. \cite{Eastwood-AMS} and more recent \cite{Mason} for the details of the ambitwistor transform). The Penrose transform for other functions in the series (\ref{wf-decomposition}) deserves more detailed treatment.

First, observe that the $SU(4)$ representations spanned by monomials composed of odd supertwistor components are in general reducible and decompose into a sum of irreducible ones, e.g. as
\beq\label{4times4}
\eta^A\bar\zeta_B=\widetilde{\eta^A\bar\zeta_B}+\frac14\de^A_B\eta^C\bar\zeta_C,\quad\widetilde{\eta^A\bar\zeta_B}=\eta^A\bar\zeta_B-\frac14\de^A_B\eta^C\bar\zeta_C
\eeq
illustrating the rule $\mathbf4\times\mathbf{\bar4}=\mathbf{15}+\mathbf1$. Widetilde over a monomial indicates that its $SU(4)$ traceless part is taken. Then multiplying (\ref{4times4}) by the function $a_{(0,0)}{}^B_A(Z,\bar W)$ allows to represent corresponding term in (\ref{wf-decomposition}) as the sum of two contributions that transform irreducibly under $SU(4)$
\beq\label{a00-irreps}
a_{(0,0)}{}^B_A(Z,\bar W)\eta^A\bar\zeta_B=\tilde a_{(0,0)}{}^B_A(Z,\bar W)\widetilde{\eta^A\bar\zeta_B}+\frac14a_{(0,0)}{}^B_B(Z,\bar W)\eta^A\bar\zeta_A.
\eeq
In the first summand tilde over $a_{(0,0)}{}^B_A$ denotes its traceless part and in the second summand ambitwistor constraint (\ref{mult-delta-function}) can be used to replace $\eta^A\bar\zeta_A$ by $-Z^{\bs{\al}}\bar W_{\bs{\al}}$. The Penrose transform of $\tilde a_{(0,0)}{}^B_A(Z,\bar W)$ yields precisely 15 vector fields $\tilde a_{\al\dot\al}{}^B_A(x)$ of $N=4$ conformal supergravity multiplet and that of $Z^{\bs{\al}}\bar W_{\bs{\al}}\, a_{(0,0)}{}^B_B(Z,\bar W)$ gives zero. The reasoning behind the latter statement is the following. In the oscillator approach the counterpart of $Z^{\bs{\al}}\bar W_{\bs{\al}}$ is the operator $-a_\al(2)a^\al(1)+b_\al(2)b^\al(1)$. The $SU(2,2)$ part of the lowest-weight vector for the $SU(2,2)\times SU(4)$ representation associated with $a_{(0,0)}{}^B_B(Z,\bar W)$ is $a^\al(1)b_\bt(2)|0\rangle$ \cite{GMarcus} and it is annihilated by $-a_\al(2)a^\al(1)+b_\al(2)b^\al(1)$. It would be interesting to find a twistor analogue of this argument.
Other ambitwistor functions $b_{(1,-1)}{}^{[AB]}(Z,\bar W)$ and $b_{(-1,1)}{}_{[AB]}(Z,\bar W)$ that appear in the second order of the decomposition (\ref{wf-decomposition}) correspond via the Penrose transform to symmetric $SL(2,\mathbb C)$ spinor fields $b_{\dot\al(2)}{}^{[AB]}(x)$ and $b_{\al(2)[AB]}(x)$, i.e. (anti)selfdual antisymmetric rank-2 tensor fields, in the $SU(4)$ representation $\mathbf6$.

Similarly the $\mathbf6\times\mathbf6$ representation spanned by the quartic monomial $\eta^A\eta^B\bar\zeta_C\bar\zeta_D$ is known to decompose as $\mathbf6\times\mathbf6=\mathbf{20'}+\mathbf{15}+\mathbf1$
\beq\label{6times6}
\begin{array}{rl}
\eta^A\eta^B\bar\zeta_C\bar\zeta_D=&\widetilde{\eta^A\eta^B\bar\zeta_C\bar\zeta_D}-\frac12\eta^E\bar\zeta_E(\de^A_C\widetilde{\eta^B\bar\zeta_D}-\de^A_D\widetilde{\eta^B\bar\zeta_C}-\de^B_C\widetilde{\eta^A\bar\zeta_D}+\de^B_D\widetilde{\eta^A\bar\zeta_C})\\[0.2cm]
-&\frac{1}{12}(\de^A_C\de^B_D-\de^A_D\de^B_C)(\eta^E\bar\zeta_E)^2,
\end{array}
\eeq
where
\beq
\begin{array}{rl}
\widetilde{\eta^A\eta^B\bar\zeta_C\bar\zeta_D}=&\eta^A\eta^B\bar\zeta_C\bar\zeta_D+\frac12\eta^E\bar\zeta_E(\de^A_C\eta^B\bar\zeta_D-\de^A_D\eta^B\bar\zeta_C-\de^B_C\eta^A\bar\zeta_D+\de^B_D\eta^A\bar\zeta_C)\\[0.2cm]
-&\frac16(\de^A_C\de^B_D-\de^A_D\de^B_C)(\eta^E\bar\zeta_E)^2
\end{array}
\eeq is traceless and transforms according to the $\mathbf{20'}$
representation. Multiplication of (\ref{6times6}) by
$\varphi_{(-1,-1)}{}_{[AB]}^{[CD]}(Z,\bar W)$ gives
\beq\label{scalar-irreps}
\begin{array}{rl}
\varphi_{(-1,-1)}{}_{[AB]}^{[CD]}(Z,\bar W)\eta^A\eta^B\bar\zeta_C\bar\zeta_D=&\tilde\varphi_{(-1,-1)}{}_{[AB]}^{[CD]}(Z,\bar W)\widetilde{\eta^A\eta^B\bar\zeta_C\bar\zeta_D}\\[0.2cm]
-&2\tilde\varphi_{(-1,-1)}{}_{CA}^{CB}(Z,\bar W)\eta^D\bar\zeta_D\widetilde{\eta^A\bar\zeta_B}\\[0.2cm]
-&\frac16\varphi_{(-1,-1)}{}_{AB}^{AB}(Z,\bar W)(\eta^C\bar\zeta_C)^2,
\end{array}
\eeq
where $\tilde\varphi_{(-1,-1)}{}_{[AB]}^{[CD]}(Z,\bar W)$ and $\tilde\varphi_{(-1,-1)}{}_{CA}^{CB}(Z,\bar W)$ are traceless.
The Penrose transform of $\tilde\varphi_{(-1,-1)}{}_{[AB]}^{[CD]}(Z,\bar W)$ produces 20 scalar fields $\tilde\varphi_{[AB]}^{[CD]}(x)$ from the $N=4$ conformal supergravity multiplet, whereas the Penrose transform of other twistor functions in (\ref{scalar-irreps}) should give zero by the argument similar to that of the previous paragraph.

Looking at other quartic terms in (\ref{wf-decomposition}) reveals two $SU(4)$ singlets $\varphi_{(-3,1)[ABCD]}(Z,\bar W)$ and $\varphi_{(1,-3)}{}^{[ABCD]}(Z,\bar W)$ together with functions $\varphi_{(-2,0)}{}_{[ABC]}^D(Z,\bar W)$ and $\varphi_{(0,-2)}{}_A^{[BCD]}(Z,\bar W)$ transforming in the reducible $SU(4)$ representations that decompose as $\mathbf4\times\mathbf4=\mathbf{10}+\mathbf6$ and $\mathbf{\bar4}\times\mathbf{\bar4}=\mathbf{\bar{10}}+\mathbf{6}$ so that we can write
%similarly to (\ref{scalar-irreps})
\beq
\begin{array}{rl}
\varphi_{(-2,0)}{}_{[ABC]}^D(Z,\bar W)\eta^A\eta^B\eta^C\bar\zeta_D=&\tilde\varphi_{(-2,0)}{}_{[ABC]}^D(Z,\bar W)\widetilde{\eta^A\eta^B\eta^C\bar\zeta_D}\\[0.2cm]
+&\frac32\varphi_{(-2,0)}{}_{D[AB]}^D(Z,\bar W)\eta^C\bar\zeta_C\eta^A\eta^B,\\[0.2cm]
\varphi_{(0,-2)}{}_A^{[BCD]}(Z,\bar W)\eta^A\bar\zeta_B\bar\zeta_C\bar\zeta_D=&\tilde\varphi_{(0,-2)}{}_A^{[BCD]}(Z,\bar W)\widetilde{\eta^A\bar\zeta_B\bar\zeta_C\bar\zeta_D}\\[0.2cm]
+&\frac32\varphi_{(0,-2)}{}_D^{D[AB]}(Z,\bar W)\eta^C\bar\zeta_C\bar\zeta_A\bar\zeta_B.
\end{array}
\eeq
Monomials with widetilde
\beq
\begin{array}{c}
\widetilde{\eta^A\eta^B\eta^C\bar\zeta_D}=\eta^A\eta^B\eta^C\bar\zeta_D-\frac12\de^A_D\eta^E\bar\zeta_E\eta^B\eta^C+\frac12\de^B_D\eta^E\bar\zeta_E\eta^A\eta^C-\frac12\de^C_D\eta^E\bar\zeta_E\eta^A\eta^B,\\
\widetilde{\eta^A\bar\zeta_B\bar\zeta_C\bar\zeta_D}=\eta^A\bar\zeta_B\bar\zeta_C\bar\zeta_D-\frac12\de^A_B\eta^E\bar\zeta_E\bar\zeta_C\bar\zeta_D+\frac12\de^A_C\eta^E\bar\zeta_E\bar\zeta_B\bar\zeta_D-\frac12\de^A_D\eta^E\bar\zeta_E\bar\zeta_B\bar\zeta_C
\end{array}
\eeq
are traceless and project out $\mathbf{10}$ and $\mathbf{\bar{10}}$ representations from $\varphi_{(-2,0)}{}_{[ABC]}^D(Z,\bar W)$ and $\varphi_{(0,-2)}{}_A^{[BCD]}(Z,\bar W)$.
Comparison with the $N=4$ conformal supergravity multiplet shows that
%$SU(4)$ singlets as well as
$\tilde\varphi_{(-2,0)}{}_{[ABC]}^D(Z,\bar W)$ and $\tilde\varphi_{(0,-2)}{}_A^{[BCD]}(Z,\bar W)$
%with the tilde denoting restriction to $\mathbf{10}$ and $\mathbf{\bar10}$ representations
should produce scalar fields via the Penrose transform. This can be explained in the following way. Ambitwistor functions with the above homogeneity degrees can be obtained by applying the first-order differential operators
\beq\label{inf-twist-operators}
\bar W_{\bs{\al}}I^{\bs{\al\bt}}\,\frac{\pt}{\pt\vp{\hat Z^{\bs{\bt}}}Z^{\bs{\bt}}},\qquad Z^{\bs{\al}}I_{\bs{\al\bt}}\,\frac{\pt}{\pt
%\vp{\hat{\hat W}_{\bs{\bt}}}
\bar W_{\bs{\bt}}}
\eeq
to the $(-1,-1)$ ambitwistor functions that correspond to the scalar fields
\beq\label{twisted-scalars}
\begin{array}{c}
\varphi_{(-2,0)}{}_{[ABC]}^D(Z,\bar W)=\left(\bar WI\frac{\pt}{\pt Z}\right)\varphi_{(-1,-1)}{}_{[ABC]}^D(Z,\bar W),\\[0.2cm]
\varphi_{(0,-2)}{}_A^{[BCD]}(Z,\bar W)=\left(ZI\frac{\pt}{\pt\vp{\hat W}\bar W}\right)\varphi_{(-1,-1)}{}_A^{[BCD]}(Z,\bar W).
\end{array}
\eeq
Differential operators (\ref{inf-twist-operators})
involve infinity twistors
\beq
I^{\bs{\al}\bs{\bt}}=\left(
\begin{array}{cc}
\varepsilon^{\al\bt} &0\\
0&0
\end{array}
\right),\quad
I_{\bs{\al}\bs{\bt}}=\left(
\begin{array}{cc}
0&0\\
0&\varepsilon^{\dot\al\dot\bt}
\end{array}\right)
\eeq
and have the following form in terms of the $SL(2,\mathbb C)$ spinor parts
\beq
\bar W_{\bs{\al}}I^{\bs{\al\bt}}\frac{\pt}{\pt Z^{\bs{\bt}}}=-v^\al\frac{\pt}{\pt\mu^\al},
\quad Z^{\bs{\al}}I_{\bs{\al\bt}}\frac{\pt}{\pt\vp{\hat\hat{ W}_{\bs{\bt}}}\bar W_{\bs{\bt}}}=-\bar u^{\dot\al}\frac{\pt}{\pt\vp{hat\nu^{\dot\al}}\bar\nu^{\dot\al}}.
\eeq
In twistor theory composition of these operators is known to give the space-time d'Alambertian and plays a role in twistor description of massive fields (see, e.g. \cite{MCP}). Let us note that the action of these operators does not change the sum of homogeneity degrees of the ambitwistor function.
So the Penrose transform of $\varepsilon^{EABC}\tilde\varphi_{(-1,-1)}{}_{[ABC]}^D(Z,\bar W)$ and $\varepsilon_{EBCD}\tilde\varphi_{(-1,-1)}{}_A^{[BCD]}(Z,\bar W)$ yields scalar fields $\varphi^{(DE)}(x)$ and $\bar\varphi_{(AE)}(x)$ in $\mathbf{10}$ and $\mathbf{\bar{10}}$ representations respectively, whereas the Penrose transform of $Z^{\bs{\al}}\bar W_{\bs{\al}}\varphi_{(-2,0)}{}_{D[AB]}^D(Z,\bar W)$ and $Z^{\bs{\al}}\bar W_{\bs{\al}}\varphi_{(0,-2)}{}_D^{D[AB]}(Z,\bar W)$ should give zero. The argument generalizes one that has been used above for the $SU(2,2)$ representations corresponding to vector and scalar fields. Using the relation between the twistor components and $SU(2)\times SU(2)$ oscillators it is possible to find the $SU(2,2)$ lowest-weight vectors for the fields on the r.h.s. of (\ref{twisted-scalars}): $(b_\al(2)b_\bt(1)-b_\bt(2)b_\al(1))|0\rangle$ and $(a^\al(2)a^\bt(1)-a^\bt(2)a^\al(1))|0\rangle$ \cite{GMarcus} and to verify that they are annihilated by the operator $-a_\al(2)a^\al(1)+b_\al(2)b^\al(1)$. Note that each antisymmetrized product of bosonic raising oscillators does not change the $SU(2)$ labels of the $SU(2,2)$ lowest-weight vector on which it acts but increases the $AdS$ energy by one unit \cite{GMZ1} and so the operators (\ref{inf-twist-operators}) are the twistor counterparts of such antisymmetrized products of raising oscillators.

Analogously ambitwistor functions $\varphi_{(-3,1)[ABCD]}(Z,\bar W)$ and $\varphi_{(1,-3)}{}^{[ABCD]}(Z,\bar W)$ can be viewed as resulting upon the repeated application of the operators (\ref{inf-twist-operators}) to the functions homogeneous of degree $(-1,-1)$
\beq
\begin{array}{c}
\varphi_{(-3,1)[ABCD]}(Z,\bar W)=\left(\bar WI\frac{\pt}{\pt\vp{\hat Z}Z}\right)^2\varphi_{(-1,-1)[ABCD]}(Z,\bar W),\\[0.2cm]
\varphi_{(1,-3)}{}^{[ABCD]}(Z,\bar W)=\left(ZI\frac{\pt}{\pt\vp{\hat{\hat W}}\bar W}\right)^2\varphi_{(-1,-1)}{}^{[ABCD]}(Z,\bar W).
\end{array}
\eeq
The Penrose transform of $\varepsilon_{ABCD}\varphi_{(-1,-1)}{}^{[ABCD]}(Z,\bar W)$ and $\varepsilon^{ABCD}\varphi_{(-1,-1)[ABCD]}(Z,\bar W)$ furnishes two remaining scalar fields $\varphi(x)$ and $\bar\varphi(x)$ of the $N=4$ conformal supergravity multiplet. Respective $SU(2,2)$ lowest weight vectors are $(b_\al(2)b_\bt(1)-b_\bt(2)b_\al(1))^2|0\rangle$ and $(a^\al(2)a^\bt(1)-a^\bt(2)a^\al(1))^2|0\rangle$ \cite{GMarcus}.

Turning to the discussion of the correspondence between the fermionic fields of $N=4$ conformal supergravity and odd components in the decomposition (\ref{wf-decomposition}) we find that eight gravitini $\psi_{\al\dot\al(2)}{}^A(x)$ and $\psi_{\al(2)\dot\al A}(x)$ are obtained via the Penrose transform of $\psi_{(1,0)}{}^A(Z,\bar W)$ and $\psi_{(0,1)A}(Z,\bar W)$ \cite{U'16}.

Spin 1/2 fields are obtained by performing the Penrose transform of the cubic terms in (\ref{wf-decomposition}). To this end let us note that similarly to the monomials in (\ref{4times4}) and (\ref{6times6}), $\eta^A\eta^B\bar\zeta_C$ and $\eta^A\bar\zeta_B\bar\zeta_C$ furnish reducible representations of $SU(4)$ that decompose as $\mathbf6\times\mathbf{\bar4}=\mathbf{20}+\mathbf4$ and $\mathbf6\times\mathbf4=\mathbf{\bar{20}}+\mathbf{\bar4}$:
\beq
\begin{array}{c}
\eta^A\eta^B\bar\zeta_C=\widetilde{\eta^A\eta^B\bar\zeta_C}+\frac13\de^B_C\eta^D\bar\zeta_D\eta^A-\frac13\de^A_C\eta^D\bar\zeta_D\eta^B,\\[0.2cm]
\eta^A\bar\zeta_B\bar\zeta_C=\widetilde{\eta^A\bar\zeta_B\bar\zeta_C}+\frac13\de^A_B\eta^D\bar\zeta_D\bar\zeta_C-\frac13\de^A_C\eta^D\bar\zeta_D\bar\zeta_B,
\end{array}
\eeq
where $\widetilde{\eta^A\eta^B\bar\zeta_C}$ and $\widetilde{\eta^A\bar\zeta_B\bar\zeta_C}$ span $\mathbf{20}$ and $\mathbf{\bar{20}}$ representations. This allows to decompose respective terms in the expansion (\ref{wf-decomposition}) as
\beq\label{24-ferm}
\begin{array}{c}
\lambda_{(-1,0)}{}^C_{[AB]}(Z,\bar W)\eta^A\eta^B\bar\zeta_C=\tilde\lambda_{(-1,0)}{}^C_{[AB]}(Z,\bar W)\widetilde{\eta^A\eta^B\bar\zeta_C}-\frac23\lambda_{(-1,0)}{}^B_{[BA]}(Z,\bar W)\eta^C\bar\zeta_C\eta^A,\\[0.2cm]
\lambda_{(0,-1)}{}^{[BC]}_{A}(Z,\bar W)\eta^A\bar\zeta_B\bar\zeta_C=\tilde\lambda_{(0,-1)}{}^{[BC]}_{A}(Z,\bar W)\widetilde{\eta^A\bar\zeta_B\bar\zeta_C}+\frac23\lambda_{(0,-1)}{}^{[BA]}_{B}(Z,\bar W)\eta^C\bar\zeta_C\bar\zeta_A.
\end{array}
\eeq
Ambitwistor functions $\tilde\lambda_{(-1,0)}{}^C_{[AB]}(Z,\bar W)$ and $\tilde\lambda_{(0,-1)}{}^{[BC]}_{A}(Z,\bar W)$ give upon the Penrose transform spin 1/2 fields $\tilde\lambda_\al{}^C_{[AB]}(x)$ and $\tilde\lambda_{\dot\al}{}^{[BC]}_{A}(x)$ in $\mathbf{\bar{20}}$ and $\mathbf{20}$ representations of $SU(4)$. Penrose transform of other ambitwistor functions in (\ref{24-ferm}) should give zero by extending the argument given above for the $SU(2,2)$ representations corresponding to bosonic fields.
Remaining third order terms in (\ref{wf-decomposition}) supply additional spin 1/2 fields in a way similar to (\ref{twisted-scalars})
\beq
\begin{array}{c}
\lambda_{(-2,1)[ABC]}(Z,\bar W)=\left(\bar WI\frac{\pt}{\pt\vp{\hat Z}Z}\right)\lambda_{(-1,0)[ABC]}(Z,\bar W), \\[0.2cm]
\lambda_{(1,-2)}{}^{[ABC]}(Z,\bar W)=\left(ZI\frac{\pt}{\pt\vp{\hat{\hat W}}\bar W}\right)\lambda_{(0,-1)}{}^{[ABC]}(Z,\bar W).
\end{array}
\eeq
Then the Penrose transform of $\varepsilon^{DABC}\lambda_{(-1,0)[ABC]}(Z,\bar W)$ and $\varepsilon_{DABC}\lambda_{(0,-1)}{}^{[ABC]}(Z,\bar W)$ yields remaining fields $\lambda_\al{}^D(x)$ and $\lambda_{\dot\al D}(x)$ from the $N=4$ conformal supergravity multiplet.

Finally terms omitted in (\ref{wf-decomposition}) do not correspond to any space-time fields since their homogeneity degrees sum up to less than -2 that is the lower bound corresponding to the scalar fields.

\section{Conclusion and discussion}
\setcounter{equation}{0}

This note addressed the issue of the definition of supertwistors for the $AdS_5\times S^5$ superspace, which isometry is described by the $PSU(2,2|4)$ supergroup that is also the superconformal symmetry of $D=4$ $N=4$ Minkowski superspace. Twistor reformulation of the massless particle model on the $\coset$ supermanifold led us to consider $SU(2)$ doublets of $c$- and $a$-type $SU(2,2|4)$ supertwistors subject to seven bosonic and eight fermionic constraints
%corresponding the $psu(2|2)\oplus u(1)$ gauge symmetry
as the $AdS_5\times S^5$ supertwistors. This justifies proposed
earlier on the group-theoretical grounds definition of
$AdS_5\times S^5$ supertwistors \cite{Bars-ads5s5} and allows to
find the incidence relations with the $AdS_5\times S^5$
supercoordinates via the $\coset$ supercoset representative.

The superparticle's Lagrangian in the supertwistor formulation is
quadratic and yields only the first-class constraints that are
generators of the $psu(2|2)\oplus u(1)$ gauge symmetry. This
should facilitate transition to the quantum theory. Here we
examined quantization of the superparticle moving in the $AdS_5$
subspace of the $AdS_5\times S^5$ superspace. In this case only four
bosonic constraints associated with the $su(2)\oplus u(1)$ gauge
symmetry are imposed on the superparticle's wave function that we
chose to depend on one $c$-type supertwistor and one dual $c$-type
supertwistor. This breaks manifest $SU(2)$ symmetry but makes
contact with the ambitwistor description of the massless fields on
$AdS_5$ space-time \cite{ASW}, \cite{U'16}. The constraints imply
that the wave function is homogeneous of degree zero in its
arguments and can be expanded in the supertwistor odd components.
Then the Penrose transform of the component ambitwistor functions
gives off-shell fields of the $D=4$ $N=4$ conformal supergravity multiplet
that serve as the boundary values for on-shell fields from the
$D=5$ $N=8$ gauged supergravity multiplet in the $AdS_5$ bulk. Quantization of the
complete model with all the constraints taken into account should
give supertwistor description of the whole spectrum of $D=10$ $N=2$
chiral supergravity compactified on $AdS_5\times S^5$ space
\cite{GMarcus}, \cite{KRvN}. To make contact with the
field-theoretic description \cite{KRvN} it is necessary to work
out the details how the Penrose transform of the homogeneous
functions of $c$-type and $a$-type supertwistors produces fields in $AdS_5$ space-time.

The superparticle model discussed in this note can be straight-forwardly generalized to that of a tensionless string with the action
\beq
\begin{array}{c}
S=\int d\tau d\s\mathscr L^{AdS_5\times S^5}_{\mathrm{stwistor\: T=0\: string}}: \\[0.2cm]
\begin{array}{rl}
\mathscr L^{AdS_5\times S^5}_{\mathrm{stwistor\: T=0\: string}}=&\frac{i}{2}\,\rho^\mu\left(\bar{\mc Z}^i_{\mc A}\pt_\mu\mc Z^{\mc A}_i-\pt_\mu\bar{\mc Z}\vp{\bar{\mc Z}}^i_{\mc A}\mc Z^{\mc A}_i\right)+\frac{i}{2}\,\rho^\mu(\bar\Psi^{i'}_{\mc A}\pt_\mu\Psi^{\mc A}_{i'}-\pt_\mu\bar\Psi^{i'}_{\mc A}\Psi^{\mc A}_{i'}) \\[0.2cm]
+&a^{ij}\bar{\mc Z}_{\mc A i}\mc Z^{\mc A}_j+s^{i'j'}\bar\Psi_{\mc A i'}\Psi^{\mc A}_{j'}+t(\bar{\mc Z}_{\mc A}^{i}\mc Z^{\mc A}_{i}+\bar\Psi_{\mc A}^{i'}\Psi^{\mc A}_{i'}) \\[0.2cm]
+&i\varkappa^i_{i'}\bar\Psi^{i'}_{\mc A}\mc Z^{\mc A}_i+i\bar\varkappa^{i'}_i\bar{\mc Z}^i_{\mc A}\Psi^{\mc A}_{i'},
\end{array}
\end{array}
\eeq extending the ambitwistor string model \cite{Geyer}.
Depending on the quantization prescription and the choice of the
vacuum \cite{Tourkine}, \cite{U'17} we expect it to provide the
world-sheet CFT interpretation for the scattering amplitudes in
$D=5$ $N=8$ gauged supergravity or to produce $D=10$ $N=2B$
higher-spin massless supermultiplets compactified on $AdS_5\times
S^5$. Generalization to the tensile string model is also feasible
and should give an interesting reformulation of the $\coset$
supercoset action \cite{MT'98}, \cite{Rahmfeld}.

We anticipate that the $c$- and $a$-type supertwistors can also be used to construct the supertwistor action for $D=5$ $N=8$ gauged supergravity and possibly for the whole tower of supermultiplets arising in the compactification of $D=10$ $N=2$ chiral supergravity on $AdS_5\times S^5$ in a way similar that of \cite{BMS}, where the super-Yang-Mills action was written in the supertwistor space.

Along the same lines it is possible to consider supertwistors and supertwistor formulations for point-like and extended objects on other supersymmetric supergravity backgrounds, in particular, those with the $OSp$-type superisometries, some of which were outlined in \cite{Bars-ads5s5}.

\appendix
\setcounter{equation}{0}
\section{Details of the spinor algebra}

In this appendix various properties of  $\g-$matrices and spinors used in the main text are collected.

Chiral antisymmetric $\g$-matrices in $D=2+4$ dimensions satisfy the Clifford algebra relations
\beq\label{rho-matrices}
\rho^{\un m}_{\bs{\al\bt}}\tilde\rho^{\un n\bs{\bt\g}}+\rho^{\un n}_{\bs{\al\bt}}\tilde\rho^{\un m\bs{\bt\g}}=-2\eta^{\un{mn}}\de^{\bs{\g}}_{\bs{\al}},\quad\eta^{\un{mn}}=\mbox{diag}(-,-,+,+,+,+)
\eeq
and the Hermitian conjugation rule
\beq\label{6d-h-conjugation}
(\tilde\rho^{\un m\bs{\al\bt}})^\dagger=H_{\bs{\bt}}{}^{\bs{\g}}\rho^{\un m}_{\bs{\g\de}}H^{\bs{\de}}{}_{\bs{\al}},
\eeq
where matrices $H_{\bs{\bt}}{}^{\bs{\g}}$ and $H^{\bs{\de}}{}_{\bs{\al}}$ equal
\beq\label{h-matrices}
H_{\bs{\al}}{}^{\bs{\bt}}=i\rho^{0}_{\bs{\al\g}}\tilde\rho^{0'\bs{\g\bt}},\quad
H^{\bs{\al}}{}_{\bs{\bt}}=i\tilde\rho^{0'\bs{\al\g}}\rho^0_{\bs{\g\bt}}.
\eeq
They satisfy $H_{\bs{\al}}{}^{\bs{\bt}}H_{\bs{\bt}}{}^{\bs{\g}}=\de_{\bs{\al}}^{\bs{\g}}$, $H^{\bs{\al}}{}_{\bs{\bt}}H^{\bs{\bt}}{}_{\bs{\g}}=\de^{\bs{\al}}_{\bs{\g}}$ and are related by the transposition $H_{\bs{\al}}{}^{\bs{\bt}}=(H^{\bs{\bt}}{}_{\bs{\al}})^{\mathrm T}$.

Passing to $D=1+4$ dimensions, matrices $\rho^{0'}_{\bs{\al\bt}}$ and $\tilde\rho^{0'\bs{\al\bt}}$ are identified with the charge conjugation matrix and its inverse
\beq
\rho^{0'}_{\bs{\al\bt}}=C_{\bs{\al\bt}},\quad\tilde\rho^{0'\bs{\al\bt}}=C^{\bs{\al\bt}}:\quad C_{\bs{\al\bt}}C^{\bs{\bt\g}}=\de^{\bs{\g}}_{\bs{\al}}
\eeq
that are used to raise and lower $Spin(1,4)$ spinor indices
\beq
\lambda^{\bs{\al}}=C^{\bs{\al\bt}}\lambda_{\bs{\bt}}\quad\lambda_{\bs{\al}}=C_{\bs{\al\bt}}\lambda^{\bs{\bt}}.
\eeq
$D=1+4$ $\g$-matrices are defined as
\beq
\g^{m'}_{\bs{\al\bt}}=i\rho^{m'}_{\bs{\al\bt}},\quad\g^{m'\bs{\al\bt}}=i\tilde\rho^{m'\bs{\al\bt}}
\eeq
and satisfy
\beq
\g^{m'}{}_{\bs{\al}}{}^{\bs{\bt}}\g^{n'}{}_{\bs{\bt}}{}^{\bs{\g}}+\g^{n'}{}_{\bs{\al}}{}^{\bs{\bt}}\g^{m'}{}_{\bs{\bt}}{}^{\bs{\g}}=-2\eta^{m'n'}\de^{\bs{\g}}_{\bs{\al}},\quad
\g^{m'}{}_{\bs{\al}}{}^{\bs{\bt}}=C_{\bs{\al\g}}\g^{m'\bs{\g\bt}}=-\g^{m'}_{\bs{\al\g}}C^{\bs{\g\bt}}.
\eeq
For them the conjugation rule (\ref{6d-h-conjugation}) transforms to
\beq
(\g^{m'}{}_{\bs{\al}}{}^{\bs{\bt}})^\dagger=\g^{0}{}_{\bs{\bt}}{}^{\bs{\g}}\g^{m'}{}_{\bs{\g}}{}^{\bs{\de}}\g^{0}{}_{\bs{\de}}{}^{\bs{\al}}
\eeq
and also $(C^{\bs{\al\bt}})^\dagger=C_{\bs{\bt\al}}$.
Antisymmetrized products of $\g$-matrices
\beq\label{so14-gamma-generators}
\g^{m'n'\bs\al}{}_{\bs\bt}=-\frac{i}{4}(\g^{m'\bs\al}{}_{\bs\g}\g^{n'\bs\g}{}_{\bs\bt}-\g^{n'\bs\al}{}_{\bs\g}\g^{m'\bs\g}{}_{\bs\bt})
\eeq
provide realization of the $so(1,4)$ algebra relations
\beq\label{so14-algebra}
[\g^{m'n'},\g^{k'l'}]=i(\eta^{m'l'}\g^{n'k'}-\eta^{m'k'}\g^{n'l'}-\eta^{n'l'}\g^{m'k'}+\eta^{n'k'}\g^{m'l'}).
\eeq

To obtain matrix form of the $so(2,4)$ generators in conformal basis $D=1+4$ $\g$-matrices $\g^{m'}{}_{\bs{\al}}{}^{\bs{\bt}}$ are expressed in terms of the $SL(2,\mathbb C)$ $\s$-matrices $\s^m_{\al\dot\al}$ and $\tilde\s^{m\dot\al\al}$
\beq
\s^m_{\al\dot\al}\tilde\s^{n\dot\al\bt}+\s^n_{\al\dot\al}\tilde\s^{m\dot\al\bt}=-2\eta^{mn}\de_\al^\bt,\quad
\tilde\s^{m\dot\al\al}\s^n_{\al\dot\bt}+\tilde\s^{n\dot\al\al}\s^m_{\al\dot\bt}=-2\eta^{mn}\de_{\dot\bt}^{\dot\al}
\eeq
as
\beq
\g^{m}{}_{\bs{\al}}{}^{\bs{\bt}}=\left(
\begin{array}{cc}
0 & \s^m_{\al\dot\bt} \\
\tilde\s^{m\dot\al\bt} & 0
\end{array}
\right),\quad
\g^{5}{}_{\bs{\al}}{}^{\bs{\bt}}=i\left(
\begin{array}{cc}
\de^\bt_\al & 0 \\
0 & -\de^{\dot\al}_{\dot\bt}
\end{array}
\right).
\eeq
The charge conjugation matrices
\beq
C_{\bs{\al\bt}}=\left(
\begin{array}{cc}
-\varepsilon_{\al\bt} & 0 \\
0 & \varepsilon^{\dot\al\dot\bt}
\end{array}
\right),\quad
C^{\bs{\al\bt}}=\left(
\begin{array}{cc}
-\varepsilon^{\al\bt} & 0 \\
0 & \varepsilon_{\dot\al\dot\bt}
\end{array}
\right)
\eeq
are realized in terms of antisymmetric unit rank 2 spinors $\varepsilon_{\al\bt}$, $\varepsilon^{\al\bt}$ and c.c. ones that are used to raise and lower $SL(2,\mathbb C)$ spinor indices.\footnote{We adopt the conventions of Ref.~\cite{Wess} for the $SL(2,\mathbb C)$ spinor algebra unless otherwise stated.}
In such a realization matrices (\ref{h-matrices}) acquire the form conventional in twistor theory
\beq
H_{\bs{\al}}{}^{\bs{\bt}}=-\g^0{}_{\bs{\al}}{}^{\bs{\bt}}=\left(
\begin{array}{cc}
0 & I \\
I & 0
\end{array}
\right),\quad
H^{\bs{\al}}{}_{\bs{\bt}}=\g^{0\,\bs{\al}}{}_{\bs{\bt}}=\left(
\begin{array}{cc}
0 & I \\
I & 0
\end{array}
\right).
\eeq

To calculate the square of the particle's 5-momentum in directions tangent to $AdS_5$ it is used the completeness relation for $D=1+4$ $\g$-matrices
\beq
\g^{m'}{}_{\bs\al}{}^{\bs\bt}\g_{m'\bs\g}{}^{\bs\de}=\de_{\bs\al}^{\bs\bt}\de_{\bs\g}^{\bs\de}-2(\de_{\bs\al}^{\bs\de}\de_{\bs\g}^{\bs\bt}-C_{\bs{\al\g}}C^{\bs{\bt\de}})
\eeq
that follows from the completeness relation for $D=2+4$ $\g$-matrices
\beq
\rho^{\un m}_{\bs{\al\bt}}\tilde\rho_{\un m}^{\bs{\g\de}}=2(\de_{\bs\al}^{\bs\g}\de_{\bs\bt}^{\bs\de}-\de_{\bs\al}^{\bs\de}\de_{\bs\bt}^{\bs\g}).
\eeq

Chiral antisymmetric $\g$-matrices in $D=6$ dimensions satisfy
\beq
\rho^I_{AB}\tilde\rho^{JBC}+\rho^J_{AB}\tilde\rho^{IBC}=2\de^{IJ}\de_A^C.
\eeq
They are connected by the Hermitian conjugation $(\rho^I_{AB})^\dagger=\tilde\rho^{IBA}$.

We adopt the following definition of $\g$- and charge conjugation matrices in $D=5$ dimensions
\beq
C_{AB}=\rho^6_{AB},\quad C^{AB}=\tilde\rho^{6AB},\quad\g^{I'}_{AB}=i\rho^{I'}_{AB},\quad\g^{I'AB}=i\tilde\rho^{I'AB}.
\eeq
So that $D=5$ Clifford algebra relations read
\beq
\g^{I'}{}_A{}^B\g^{J'}{}_B{}^C+\g^{J'}{}_A{}^B\g^{I'}{}_B{}^C=2\de^{I'J'}\de_A^C,\quad\g^{I'}{}_A{}^B=C_{AD}\g^{I'DB}=-\g^{I'}_{AD}C^{DB}.
\eeq
Hermitian conjugation transforms the $D=5$ $\g$- and charge conjugation matrices as
\beq
(\g^{I'}{}_A{}^B)^\dagger=\g^{I'}{}_B{}^A,\quad(C_{AB})^\dagger=C^{BA}.
\eeq
Antisymmetrized products of $\g$-matrices
\beq
\g^{I'J'A}{}_{B}=\frac{i}{4}(\g^{I'A}{}_C\g^{J'C}{}_{B}-\g^{J'A}{}_C\g^{I'C}{}_{B})
\eeq
satisfy the commutation relations of the $so(5)$ algebra
\beq
[\g^{I'J'},\g^{K'L'}]=i(\de^{I'L'}\g^{J'K'}-\de^{I'K'}\g^{J'L'}-\de^{J'L'}\g^{I'K'}+\de^{J'K'}\g^{I'L'}).
\eeq

Completeness relation for $D=6$ $\g$-matrices
\beq
\rho^I_{AB}\tilde\rho^{ICD}=2(\de_A^D\de_B^C-\de_A^C\de_B^D)
\eeq
in terms of $D=5$ $\g$-matrices acquires the form
\beq
\g^{I'}{}_A{}^B\g^{I'}{}_C{}^D=-\de_A^B\de_C^D+2(\de_A^D\de_C^B-C_{AC}C^{BD})
\eeq
and is used to calculate the square of the particle's momentum components in directions tangent to $S^5$.

\setcounter{equation}{0}
\section{$(4|4)$ supermatrix realization of $psu(2,2|4)$ superalgebra generators}

This Appendix contains the details of the supermatrix form of the generators of $psu(2,2|4)$ superalgebra and its realization as the $D=4$ $N=4$ superconformal algebra.

$so(2,4)\sim su(2,2)$ generators can be realized by the antisymmetrized products of $D=2+4$ $\g$-matrices (\ref{rho-matrices})
\beq
\tilde\rho^{\un{mn}\bs{\al}}{}_{\bs{\bt}}=-\frac{i}{4}(\tilde\rho^{\un m\bs{\al\g}}\rho^{\un n}_{\bs{\g\bt}}-\tilde\rho^{\un n\bs{\al\g}}\rho^{\un m}_{\bs{\g\bt}}),
\eeq
which commutator equals
\beq\label{so24-algebra}
[\tilde\rho^{\un{mn}},\tilde\rho^{\un{kl}}]=i(\eta^{\un{ml}}\tilde\rho^{\un{nk}}-\eta^{\un{mk}}\tilde\rho^{\un{nl}}-\eta^{\un{nl}}\tilde\rho^{\un{mk}}+\eta^{\un{nk}}\tilde\rho^{\un{ml}}).
\eeq
Retaining $so(1,4)$ or $so(1,3)$ covariance the $so(2,4)$ algebra relations (\ref{so24-algebra}) can be cast in the form of $ads_5$ or $conf_4$ algebras.
The $ads_5$ algebra relations are obtained by splitting $so(2,4)$ generators into $\tilde\rho^{0'm'\bs\al}{}_{\bs\bt}=-\frac12\g^{m'\bs\al}{}_{\bs\bt}$ and $\tilde\rho^{m'n'\bs\al}{}_{\bs\bt}=\g^{m'n'\bs\al}{}_{\bs\bt}$. Then one derives from (\ref{so24-algebra})
\beq
\begin{array}{c}
[\g^{m'},\g^{n'}]=4i\g^{m'n'},\quad
[\g^{m'n'},\g^{k'}]=i(\eta^{n'k'}\g^{m'}-\eta^{m'k'}\g^{n'})
%,\\[0.2cm]
%[\tilde\rho^{m'n'},\tilde\rho^{k'l'}]
%=i(\eta^{m'l'}\tilde\rho^{n'k'}-\eta^{m'k'}\tilde\rho^{n'l'}-\eta^{n'l'}\tilde\rho^{m'k'}+\eta^{n'k'}\tilde\rho^{m'l'}).
\end{array}
\eeq
and relations (\ref{so14-algebra}). Generators of the $conf_4$ algebra are defined as
\beq\label{conf-matrix-generators}
\begin{array}{cc}
D^{\bs{\al}}{}_{\bs{\bt}}=-\tilde\rho^{0'5\bs{\al}}{}_{\bs{\bt}}=\frac{i}{2}\left(
\begin{array}{cc}
-\de^\al_\bt & 0 \\[0.2cm]
0 & \de_{\dot\al}^{\dot\bt}
\end{array}
\right),\quad
P^{m\,\bs\al}{}_{\bs\bt}=\tilde\rho^{0'm\bs{\al}}{}_{\bs{\bt}}+\tilde\rho^{5m\bs{\al}}{}_{\bs{\bt}}=\left(
\begin{array}{cc}
0 & \tilde\s_m^{\dot\bt\al} \\[0.2cm]
0 & 0
\end{array}
\right),
\\[0.3cm]
K^{m\,\bs\al}{}_{\bs\bt}=\tilde\rho^{0'm\bs{\al}}{}_{\bs{\bt}}-\tilde\rho^{5m\bs{\al}}{}_{\bs{\bt}}=\left(
\begin{array}{cc}
0 & 0 \\
\s^m_{\bt\dot\al} & 0
\end{array}
\right),\quad
M^{mn\,\bs\al}{}_{\bs\bt}=\tilde\rho^{mn\bs\al}{}_{\bs\bt}=i\left(
\begin{array}{cc}
\s^{mn}{}_\bt{}^\al & 0 \\
0 & \tilde\s^{mn\dot\bt}{}_{\dot\al}
\end{array}
\right),
\end{array}
\eeq
where $\s^{mn}{}_\bt{}^\al=\frac14(\s^m_{\bt\dot\bt}\tilde\s^{n\dot\bt\al}-\s^n_{\bt\dot\bt}\tilde\s^{m\dot\bt\al})$ and $\tilde\s^{mn\dot\bt}{}_{\dot\al}=\frac14(\tilde\s^{m\dot\bt\bt}\s^n_{\bt\dot\al}-\tilde\s^{n\dot\bt\bt}\s^m_{\bt\dot\al})$. Then the $conf_4$ algebra relations read
\beq\label{conf-algebra}
\begin{array}{c}
[D,P^m]=-iP^m,\quad [D,K^m]=iK^m,\quad [K^m,P^n]=2i(\eta^{mn}D+M^{mn}),\\[0.2cm]
[M^{mn},P^k]=i(\eta^{nk}P^m-\eta^{mk}P^n),\quad [M^{mn},K^k]=i(\eta^{nk}K^m-\eta^{mk}K^n),\\[0.2cm]
[M^{mn},M^{kl}]=i(\eta^{ml}M^{nk}-\eta^{mk}M^{nl}-\eta^{nl}M^{mk}+\eta^{nk}M^{ml}).
\end{array}
\eeq

Similarly to the case of $so(2,4)$ algebra the generators of
$so(6)\sim su(4)$ algebra can be realized by the antisymmetrized
products of $D=6$ chiral $\g$-matrices \beq
\tilde\rho^{IJA}{}_B=\frac{i}{4}(\tilde\rho^{IAC}\rho^J_{CB}-\tilde\rho^{JAC}\rho^I_{CB}).
\eeq Their commutator equals \beq\label{so6-algebra}
[\tilde\rho^{IJ},\tilde\rho^{KL}]=i(\de^{IL}\tilde\rho^{JK}-\de^{IK}\tilde\rho^{JL}-\de^{JL}\tilde\rho^{IK}+\de^{JK}\tilde\rho^{IL}).
\eeq Introducing the $s^5$ algebra generators
\beq\label{s5-matrix-generators}
P^{I'A}{}_B=2\tilde\rho^{6I'A}{}_{B}=\g^{I'A}{}_B,\quad
M^{I'J'A}{}_B=\tilde\rho^{I'J'A}{}_{B}=\g^{I'J'A}{}_B \eeq
relations (\ref{so6-algebra}) can be written in the form
\beq\label{su4-algebra}
\begin{array}{c}
[P^{I'},P^{J'}]=-4iM^{I'J'},\quad [P^{I'},M^{K'L'}]=i(\de^{I'K'}P^{L'}-\de^{I'L'}P^{K'}), \\[0.2cm]
[M^{I'J'},M^{K'L'}]=i(\de^{I'L'}M^{J'K'}-\de^{I'K'}M^{J'L'}-\de^{J'L'}M^{I'K'}+\de^{J'K'}M^{I'L'}).
\end{array}
\eeq

Generators of the Poincare supersymmetry are given by the supermatrices
\beq\label{spoincare-matrix-generators}
Q_\al^{A\,\mc B}{}_{\mc C}=\left(
\begin{array}{ccc}
0 & 0 & 2\de_\al^\bt\de^A_C \\[0.2cm]
0 & 0 & 0 \\[0.2cm]
0 & 0 & 0
\end{array}
\right),\quad
\bar Q_{\dot\al A}{}^{\mc B}{}_{\mc C}=\left(
\begin{array}{ccc}
0 & 0 & 0 \\[0.2cm]
0 & 0 & 0 \\[0.2cm]
0 & 2\de_{\dot\al}^{\dot\g}\de_A^B & 0
\end{array}
\right)
\eeq
and generators of the conformal supersymmetry equal
\beq\label{sconf-matrix-generators}
S_A^{\al\,\mc B}{}_{\mc C}=\left(
\begin{array}{ccc}
0 & 0 & 0 \\[0.2cm]
0 & 0 & 0 \\[0.2cm]
2\de^\al_\g\de_A^B & 0 & 0
\end{array}
\right),\quad
\bar S^{\dot\al A\,\mc B}{}_{\mc C}=\left(
\begin{array}{ccc}
0 & 0 & 0 \\[0.2cm]
0 & 0 & 2\de^{\dot\al}_{\dot\bt}\de^A_C \\[0.2cm]
0 & 0 & 0
\end{array}
\right).
\eeq
The reasoning behind this definition is that the $(4|4)\times(4|4)$ supermatrix $g^{\mc A}{}_{\mc B}$:
\beq
g^{\mc A}{}_{\mc B}=\left(
\begin{array}{cc}
g^{\bs\al}{}_{\bs\bt} & g^{\bs\al}{}_B \\[0.2cm]
g^A{}_{\bs\bt} & g^A{}_B
\end{array}
\right)\in psu(2,2|4)
\eeq
with the $4\times4$ blocks
\beq
\begin{array}{c}
g^{\bs\al}{}_{\bs\bt}=X^{\un{mn}}\tilde\rho_{\un{mn}}{}^{\bs\al}{}_{\bs\bt}\in su(2,2),\quad
g^A{}_B=Y^{IJ}\tilde\rho^{IJA}{}_B\in su(4), \\[0.2cm]
g^{\bs\al}{}_B=\theta^\lambda_DQ_\lambda^{D\,\bs\al}{}_{B}+\bar\eta_{\dot\lambda D}\bar S^{\dot\lambda D\,\bs\al}{}_B,\quad
g^A{}_{\bs\bt}=\bar\theta^{\dot\lambda D}\bar Q_{\dot\lambda D}{}^A{}_{\bs\bt}+\eta^D_\lambda S_D^{\lambda\, A}{}_{\bs\bt}
\end{array}
\eeq
satisfies the Hermiticity condition
\beq
g^{\mc A}{}_{\mc B}=\mc H^{\mc A}{}_{\mc C}(g^{\mc D}{}_{\mc C})^\dagger\mc H^{\mc D}{}_{\mc B},
\eeq
where
\beq
(g^{\mc A}{}_{\mc B})^\dagger=\left(
\begin{array}{cc}
(g^{\bs\al}{}_{\bs\bt})^\dagger & (g^A{}_{\bs\bt})^\dagger \\[0.2cm]
(g^{\bs\al}{}_B)^\dagger & (g^A{}_B)^\dagger
\end{array}
\right)
\eeq
and
\beq
\mc H^{\mc A}{}_{\mc B}=\left(
\begin{array}{cc}
H^{\bs{\al}}{}_{\bs\bt} & 0 \\[0.2cm]
0 & \de^A_B
\end{array}\right).
\eeq
Thus explicit supermatrix form of the general $psu(2,2|4)$ element in the $SL(2,\mathbb C)$ notation is
\beq
\begin{array}{c}
g^{\mc A}{}_{\mc B}=\left(
\begin{array}{ccc}
-i(X^{0'5}\de^\al_\bt-X^{mn}\s_{mn\bt}{}^\al) & (X^{5m}-X^{0'm})\tilde\s_m^{\dot\bt\al} & 2\theta^\al_B \\[0.2cm]
 -(X^{0'm}+X^{5m})\s_{m\bt\dot\al} & i(X^{0'5}\de_{\dot\al}^{\dot\bt}+X^{mn}\tilde\s_{mn}{}^{\dot\bt}{}_{\dot\al}) & 2\bar\eta_{\dot\al B} \\[0.2cm]
2\eta^A_\bt & 2\bar\theta^{\dot\bt A} & Y^{IJ}\tilde\rho^{IJA}{}_B
\end{array}
\right).
\end{array}
\eeq

Using the above introduced matrix form of the $psu(2,2|4)$ generators (\ref{conf-matrix-generators}), (\ref{s5-matrix-generators}), (\ref{spoincare-matrix-generators}) and (\ref{sconf-matrix-generators}) it is possible to derive (anti)commutation relations of the $psu(2,2|4)$ superalgebra realized as the $D=4$ $N=4$ superconformal algebra. Commutation relations of the $su(2,2)\oplus su(4)$ bosonic subalgebra are given in (\ref{conf-algebra}) and (\ref{su4-algebra}). Non-zero anticommutators of odd generators equal
\beq
\begin{array}{c}
\{Q^C_{\lambda},\bar Q_{\dot\lambda D}\}=-2\de^C_D\s^m_{\lambda\dot\lambda}P_m,\quad\{S^\lambda_C,\bar S^{\dot\lambda D}\}=-2\de^D_C\tilde\s^{m\dot\lambda\lambda}K_m, \\[0.2cm]
\{Q^C_\lambda,S^\rho_D\}=2i\de^C_D(\de^\rho_\lambda D+\s^{mn}{}_\lambda{}^\rho M_{mn})+\de^\rho_\lambda(\g^{I'C}{}_DP^{I'}+2\g^{I'J'C}{}_DM^{I'J'}), \\[0.2cm]
\{\bar Q_{\dot\lambda C},\bar S^{\dot\rho D}\}=-2i\de^D_C(\de^{\dot\rho}_{\dot\lambda}D-\tilde\s^{mn\dot\rho}{}_{\dot\lambda}M_{mn})+\de^{\dot\rho}_{\dot\lambda}(\g^{I'D}{}_CP^{I'}+2\g^{I'J'D}{}_CM^{I'J'})
\end{array}
\eeq
and commutators of odd and even generators read
\beq
\begin{array}{c}
[D,Q^C_\lambda]=-\frac{i}{2}Q^C_\lambda,\quad [D,\bar Q_{\dot\lambda C}]=-\frac{i}{2}\bar Q_{\dot\lambda C},\quad
[D,S^\lambda_C]=\frac{i}{2}S^\lambda_C,\quad [D,\bar S^{\dot\lambda C}]=\frac{i}{2}\bar S^{\dot\lambda C}, \\[0.2cm]
[P_m,S^\lambda_C]=-\tilde\s_m^{\dot\lambda\lambda}\bar Q_{\dot\lambda C},\quad [P_m,\bar S^{\dot\lambda C}]=\tilde\s^{\dot\lambda\lambda}_mQ^C_\lambda, \\[0.2cm]
[K_m,Q^C_\lambda]=\s_{m\lambda\dot\lambda}\bar S^{\dot\lambda C},\quad
[K_m,\bar Q_{\dot\lambda C}]=-\s_{m\lambda\dot\lambda}S^\lambda_C, \\[0.2cm]
[M_{mn},Q^C_\lambda]=i\s_{mn\lambda}{}^\rho Q^C_\rho,\quad [M_{mn},\bar Q_{\dot\lambda C}]=-i\tilde\s_{mn}{}^{\dot\rho}{}_{\dot\lambda}\bar Q_{\dot\rho C}, \\[0.2cm]
[M_{mn},S^\lambda_C]=-i\s_{mn\rho}{}^\lambda S^\rho_C,\quad [M_{mn},\bar S^{\dot\lambda C}]=i\tilde\s_{mn}{}^{\dot\lambda}{}_{\dot\rho}\bar S^{\dot\rho C}, \\[0.2cm]
[P^{I'},Q^C_\lambda]=-\g^{I'C}{}_DQ^D_\lambda,\quad [P^{I'},\bar Q_{\dot\lambda C}]=\g^{I'D}{}_C\bar Q_{\dot\lambda D}, \\[0.2cm]
[P^{I'},S^\lambda_C]=\g^{I'D}{}_CS^\lambda_D,\quad [P^{I'},\bar S^{\dot\lambda C}]=-\g^{I'C}{}_{D}\bar S^{\dot\lambda D}, \\[0.2cm]
[M^{I'J'},Q^C_\lambda]=-\g^{I'J'C}{}_DQ^D_\lambda,\quad [M^{I'J'},\bar Q_{\dot\lambda C}]=\g^{I'J'D}{}_C\bar Q_{\dot\lambda D}, \\[0.2cm]
[M^{I'J'},S^\lambda_C]=\g^{I'J'D}{}_CS^\lambda_D,\quad [M^{I'J'},\bar S^{\dot\lambda C}]=-\g^{I'J'C}{}_{D}\bar S^{\dot\lambda D}.
\end{array}
\eeq


\begin{thebibliography}{99}
\bibitem{Penrose}
R.~Penrose, Twistor algebra, J. Math. Phys. \textbf{8} (1967) 345.
\bibitem{Ferber}
A.~Ferber, Supertwistors and conformal supersymmetry, Nucl. Phys. \textbf{B132} (1978) 55.
\bibitem{theta-twistor}
A.A.~Zheltukhin, Unification of twistors and Ramond
vectors, Phys. Lett. {\bf B658} (2007) 82, \href{http://arxiv.org/abs/0707.3453}{arXiv:0707.3453 [hep-th]}.


\bibitem{Shirafuji}
T.~Shirafuji, Lagrangian mechanics of massless particles with spin, Prog. Theor. Phys. \textbf{70} (1983) 18.
\bibitem{bbcl}
A.~Bengtsson, I.~Bengtsson, M.~Cederwall and N.~Linden,
Particles, superparticles and twistors, Phys. Rev. {\bf
D36} (1987) 1766.
\bibitem{Eisenberg}
Y.~Eisenberg and S.~Solomon, The twistor geometry of the
covariantly quantized Brink-Schwarz superparticle, Nucl. Phys.
{\bf B309} (1988) 709.
\bibitem{Plyushchay}
M.~Plyushchay, Covariant quantization of massless
superparticle in four-dimensional space-time: twistor approach,
Mod. Phys. Lett. {\bf A4} (1989) 1827. \\
M.~Plyushchay, Lagrangian formulation for the massless (super)particle in the (super)twistor
approach, Phys. Lett. {\bf B240} (1990) 133.
\bibitem{Gumenchuk}
A.~Gumenchuk and D.~Sorokin, Relativistic superparticle
dynamics and twistor correspondence, Sov. J. Nucl. Phys. {\bf51}
(1990) 350 [Yad. Fiz. \textbf{51} (1990) 549].
\bibitem{DVVolkov}
I.A.~Bandos, A.~Nurmagambetov, D.~Sorokin and D.V.~Volkov, Twistor-like superparticles revisited, Class. Quantum Grav.
\textbf{12} (1995) 1881, \href{http://arxiv.org/abs/hep-th/9502143}{arXiv:hep-th/9502143}.

\bibitem{Zima}
S.~Fedoruk and V.G.~Zima, Bitwistor formulation of massive spinning particle, J. of Kharkiv University \textbf{585} (2003) 39, \href{http://arxiv.org/abs/hep-th/0308154}{arXiv:hep-th/0308154}.
\bibitem{Bette}
A.~Bette, J.A.~de Azcarraga, J.~Lukierski and C.~Miquel-Espanya,  Massive relativistic particle model with spin and electric charge from two twistor dynamics, Phys.\ Lett.\  {\bf B595} (2004) 491, \href{http://arxiv.org/abs/hep-th/0405166}{arXiv:hep-th/0405166}.
\bibitem{Picon1}
I.~Bars and M.~Picon, Single twistor description of
massless, massive, AdS, and other interacting particles, Phys.
Rev. {\bf D73} (2006) 064002, \href{http://arxiv.org/abs/hep-th/0512091}{arXiv:hep-th/0512091}.
\bibitem{dAIL}
J.A.~de Azcarraga, J.M.~Izquierdo and J.~Lukierski, Supertwistors, massive superparticles and $\kappa$-symmetry, JHEP {\bf 0901} (2009) 041, \href{http://arxiv.org/abs/0808.2155}{arXiv:0808.2155 [hep-th]}.
\bibitem{Mezincescu'13}
L.~Mezincescu, A.J.~Routh and P.K.~Townsend, Supertwistors and massive particles, Ann. Phys.\  {\bf 346} (2014) 66, \href{http://arxiv.org/abs/1312.2768}{arXiv:1312.2768[hep-th]}.

\bibitem{Ilyenko}
K.~Ilyenko, Twistor variational principle for null strings,
Nucl. Phys. B (Proc. Suppl.) {\bf102-103} (2001) 83, \href{http://arxiv.org/abs/hep-th/0104117}{arXiv:hep-th/0104117}. \\
K.~Ilyenko, Twistor representation of null two
surfaces, J. Math. Phys. {\bf43} (2002) 4770, \href{http://arxiv.org/abs/hep-th/0109036}{arXiv:hep-th/0109036}.
\bibitem{U-CQG06}
D.V.~Uvarov, (Super)twistors and (super)strings, Class. Quantum Grav. {\bf23} (2006) 2711, \href{http://arxiv.org/abs/hep-th/0601149}{arXiv:hep-th/0601149}.
\bibitem{BdAM}
I.A.~Bandos, J.A.~de Azcarraga and C.~Miquel-Espanya, Superspace
formulations of the (super)twistor string, JHEP {\bf0607} (2006)
005,
\href{http://arxiv.org/abs/hep-th/0604037}{arXiv:hep-th/0604037}.
\bibitem{FL06}
S.~Fedoruk and J.~Lukierski, Twistorial versus space-time formulations: unification of various string models, Phys.\ Rev.\ {\bf D75} (2007) 026004, \href{http://arxiv.org/abs/hep-th/0606245}{arXiv:hep-th/0606245}. \\
S.~Fedoruk and J.~Lukierski, Purely twistorial string with canonical twistor field quantization, Phys.\ Rev.\ {\bf D79} (2009) 066006, \href{http://arxiv.org/abs/0811.3353}{arXiv:0811.3353 [hep-th]}.
\bibitem{FL07}
S.~Fedoruk and J.~Lukierski, Two-twistor description of membrane, Phys.\ Rev.\ {\bf D76} (2007) 066005, \href{http://arxiv.org/abs/0706.2129}{arXiv:0706.2129 [hep-th]}.

\bibitem{bc}
I.~Bengtsson and M.~Cederwall, Particles, twistors and the
division algebras, Nucl. Phys. {\bf B302} (1988) 81.
\bibitem{Berkovits'90}
N.~Berkovits, A supertwistor description of the massless superparticle in ten-dimensional superspace, Phys.\ Lett.\ {\bf B247} (1990) 45; Nucl.\ Phys.\ {\bf B350} (1991) 193.
\bibitem{Howe-West}
P.S.~Howe and P.C.~West, The conformal group, point particles and twistors, Int. J.
Mod. Phys. A7 (1992) 6639.
\bibitem{BLS}
I.~Bandos and J.~Lukierski, Tensorial central charges and
new superparticle model with fundamental spinor coordinates, Mod.
Phys. Lett. {\bf A14} (1999) 1257, \href{http://arxiv.org/abs/hep-th/9811022}{arXiv:hep-th/9811022}. \\
I.A. Bandos, J. Lukierski and D.P. Sorokin, Superparticle
models with tensorial central charges, Phys. Rev. {\bf D61} (2000)
045002, \href{http://arxiv.org/abs/hep-th/9904109}{arXiv:hep-th/9904109}.
\bibitem{BdAPV}
I.A.~Bandos, J.A.~de Azcarraga, M.~Picon and O.~Varela, $D=11$
superstring model with 30 $\kappa-$symmetries and $\frac{30}{32}$
BPS states in an extended superspace, Phys. Rev. {\bf D69} (2004)
085007, \href{http://arxiv.org/abs/hep-th/0307106}{arXiv:hep-th/0307106}.
\bibitem{Picon2}
I.~Bars and M.~Picon, Twistor transform in $d$ dimensions and a unifying role for
twistors, Phys. Rev. {\bf D73} (2006) 064033, \href{http://arxiv.org/abs/hep-th/0512348}{arXiv:hep-th/0512348}.
\bibitem{U-CQG07}
D.V.~Uvarov, Supertwistor formulation for higher dimensional superstrings, Class. Quantum Grav. {\bf24} (2007) 5383, \href{http://arxiv.org/abs/hep-th/0703051}{arXiv:hep-th/0703051}.
\bibitem{Routh}
A.J.~Routh and P.K.~Townsend, Twistor form of massive $6D$ superparticle, J.\ Phys. A: Math. Theor. {\bf 49} (2016) 025402, \href{http://arxiv.org/abs/1507.05218}{arXiv:1507.05218 [hep-th]}.
\bibitem{Bandos'14}
I.~Bandos, Twistor/ambitwistor strings and null-superstrings in space-time of $D=4,10$ and $11$ dimensions, JHEP \textbf{1409} (2014) 086, \href{http://arxiv.org/abs/1404.1299}{arXiv:1404.1299 [hep-th]}.

\bibitem{Cherkis}
N.~Berkovits and S.A.~Cherkis, Higher-dimensional twistor transform using pure spinors, JHEP \textbf{0412} (2004) 049, \href{http://arxiv.org/abs/hep-th/0409243}{arXiv:hep-th/0409243}.
\bibitem{Berkovits'10}
N.~Berkovits, Ten-dimensional super-twistors and super-Yang-Mills, JHEP \textbf{1004} (2010) 067, \href{http://arxiv.org/abs/0910.1684}{arXiv:0910.1684 [hep-th]}.
\bibitem{Elvang}
H.~Elvang and Yu-tin Huang, Scattering Amplitudes in Gauge Theory and Gravity, Cambridge University Press, 2015, \href{http://arxiv.org/abs/1308.1697}{arXiv:1308.1697 [hep-th]}.

\bibitem{CRZ}
P.~Claus, J.~Rahmfeld and Y.~Zunger, A simple particle action from a twistor parametrization of $AdS_5$, Phys.~Lett. \textbf{B466} (1999) 181, \href{http://arxiv.org/abs/hep-th/9906118}{arXiv:hep-th/9906118}.
\bibitem{ASW}
T.~Adamo, D.~Skinner and J.~Williams, Twistor methods for $AdS_5$, JHEP \textbf{1608} (2016) 167, \href{http://arxiv.org/abs/1607.03763}{arXiv:1607.03763 [hep-th]}.
\bibitem{Cederwall}
M.~Cederwall, Geometric construction of $AdS$ twistors, Phys.\ Lett.\ {\bf B483} (2000) 257, \href{http://arxiv.org/abs/hep-th/0002216}{arXiv:hep-th/0002216}.\\
M.~Cederwall, AdS twistors for higher spin theory,  AIP Conf.\ Proc.\  {\bf 767} (2005) 96, \href{http://arxiv.org/abs/hep-th/0412222}{arXiv:hep-th/0412222}.



\bibitem{ABGPKT16}
A.S.~Arvanitakis, A.E.~Barns-Graham and P.K.~Townsend, Anti-de Sitter particles and manifest (super)isometries, Phys. Rev. Lett. \textbf{118} (2017) 141601, \href{http://arxiv.org/abs/1608.04380}{arXiv:1608.04380 [hep-th]}.
\bibitem{ABGPKT17}
A.S.~Arvanitakis, A.E.~Barns-Graham and P.K.~Townsend, Twistor description of spinning particles in $AdS$, JHEP \textbf{1801} (2018) 059, \href{http://arxiv.org/abs/1710.09557}{arXiv:1710.09557 [hep-th]}.

\bibitem{MT'98}
R.R.~Metsaev and A.A.~Tseytlin, Type IIB superstring action in
$AdS_5\times S^5$ background, Nucl.\ Phys.\ \textbf{B533} (1998)
109, \href{http://arxiv.org/abs/hep-th/9805028}{arXiv:hep-th/9805028}.
\bibitem{Rahmfeld}
R.~Kallosh, J.~Rahmfeld and A.~Rajaraman, Near horizon
superspace, JHEP \textbf{9809} (1998) 002, \href{http://arxiv.org/abs/hep-th/9805217}{arXiv:hep-th/9805217}.
\bibitem{BILS}
I.~Bandos, E.~Ivanov, J.~Lukierski and D.~Sorokin, On the superconformal flatness of $AdS$ superspaces, JHEP \textbf{0206} (2002) 040, \href{http://arxiv.org/abs/hep-th/0205104}{arXiv:hep-th/0205104}.

\bibitem{Bars-ads5s5}
I.~Bars, Twistors and 2T-physics, AIP Conf. Proc. \textbf{767} (2005) 3, \href{http://arxiv.org/abs/hep-th/0502065}{arXiv:hep-th/0502065}. \\
I.~Bars, Lectures on twistors, \href{http://arxiv.org/abs/hep-th/0601091}{arXiv:hep-th/0601091}.
\bibitem{Siegel-twistor}
W.~Siegel, AdS/CFT in superspace, \href{http://arxiv.org/abs/1005.2317}{arXiv:1005.2317 [hep-th]}. \\
W.~Siegel, Embedding versus $6D$ twistors, \href{http://arxiv.org/abs/1204.5679}{arXiv:1204.5679 [hep-th]}.
\bibitem{Buchbinder}
I.L.~Buchbinder and S.M.~Kuzenko, Ideas and Methods of Supersymmetry and Supergravity or A Walk Through Superspace, IOP Publishing, 1998.
\bibitem{Bars-twistor-string}
I.~Bars, Twistor superstring in 2T-physics, Phys. Rev. \textbf{D70} (2004) 104022, \href{http://arxiv.org/abs/hep-th/0407239}{arXiv:hep-th/0407239}.
\bibitem{Uf}
D.V.~Uvarov, in preparation.

\bibitem{Galperin1}
F.~Delduc, A.S.~Galperin and E.~Sokatchev, Lorentz-harmonic (super)fields and (super)particles, Nucl. Phys. \textbf{B368} (1992) 143.
\bibitem{Galperin2}
A.S.~Galperin, P.S.~Howe and K.S.~Stelle, The superparticle and the Lorentz group, Nucl. Phys. \textbf{B368} (1992) 248, \href{http://arxiv.org/abs/hep-th/9201020}{arXiv:hep-th/9201020}.
\bibitem{BZh}
I.A.~Bandos and A.A.~Zheltukhin, Spinor Cartan moving $n-$hedron, Lorentz-harmonic formulations of superstrings and $\kappa-$symmetry, JETP Lett. \textbf{54} (1991) 421 [Pis'ma v ZhETF \textbf{54} (1991) 421]. \\
I.A.~Bandos and A.A.~Zheltukhin, Green-Schwarz superstrings in spinor moving frame formalism, Phys. Lett. \textbf{B288} (1992) 77.

\bibitem{GMarcus}
M.~Gunaydin and N.~Marcus, The spectrum of the $S^5$ compactification of the chiral $N=2$, $D=10$ supergravity and the unitary multiplets of $U(2,2|4)$,  Class. Quantum Grav. \textbf{2} (1985) L11.
\bibitem{KRvN}
H.J.~Kim, L.J.~Romans and P.~van Nieuwenhuizen, Mass spectrum of chiral ten-dimensional $N=2$ supergravity on $S^5$, Phys. Rev. \textbf{D32} (1985) 389.



\bibitem{MTT}
R.R.~Metsaev, C.B.~Thorn and A.A.~Tseytlin, Light cone superstring in AdS space-time,  Nucl.\ Phys.\ \textbf{B596} (2001) 151, \href{http://arxiv.org/abs/hep-th/0009171}{arXiv:hep-th/0009171}.
\bibitem{Bars'02}
I.~Bars, Hidden twelve-dimensional structures in $AdS_5\times S^5$ and $M^4\times R^6$ supergravities, Phys.\ Rev.\ {\bf D66} (2002) 105024, \href{http://arxiv.org/abs/hep-th/0208012}{arXiv:hep-th/0208012}.
\bibitem{Horigane}
T.~Horigane and Y.~Kazama, Exact quantization of a superparticle in $AdS_5\times S^5$, Phys.\ Rev.\ {\bf D81} (2010) 045004, \href{http://arxiv.org/abs/0912.1166}{arXiv:0912.1166 [hep-th]}.
\bibitem{Siegel-superparticle}
W.~Siegel, Spacecone quantization of $AdS$ superparticle, \href{http://arxiv.org/abs/1005.5049}{arXiv:1005.5049 [hep-th]}.
\bibitem{Heinze}
M.~Heinze, Spectrum and quantum symmetries of the $AdS_5\times S^5$ superstring, \href{http://arxiv.org/abs/1507.03005}{arXiv:1507.03005 [hep-th]}.


\bibitem{CKR}
P.~Claus, R.~Kallosh and J.~Rahmfeld, BRST quantization of a particle in $AdS_5$, Phys. Lett. \textbf{B462} (1999) 285, \href{http://arxiv.org/abs/hep-th/9906195}{arXiv:hep-th/9906195}.
\bibitem{U'15}
D.V.~Uvarov, Spinor description of $D=5$ massless low-spin gauge fields, Class. Quantum Grav. \textbf{33} (2016) 135010, \href{http://arxiv.org/abs/1506.01881}{arXiv:1506.01881 [hep-th]}.
\bibitem{MTlc}
R.~R.~Metsaev and A.~A.~Tseytlin, Superstring action in $AdS_5\times S^5$ : $\kappa-$symmetry light cone gauge, Phys. Rev. \textbf{D63} (2001) 046002,  \href{http://arxiv.org/abs/hep-th/0007036}{arXiv:hep-th/0007036}.
\bibitem{Metsaev'01}
R.~R.~Metsaev, On manifest $SU(4)$ invariant superstring action in $AdS_5\times S^5$, Class. Quantum Grav. \textbf{18} (2001) 1245, \href{http://arxiv.org/abs/hep-th/0012026}{arXiv:hep-th/0012026}.

\bibitem{GMZ1}
M.~Gunaydin, D.~Minic and M.~Zagermann, Novel supermultiplets of $SU(2,2|4)$ and the $AdS_5/CFT_4$ duality, Nucl. Phys. \textbf{B544} (1999) 737, \href{http://arxiv.org/abs/hep-th/9810226}{arXiv:hep-th/9810226}.
\bibitem{CGKRZ}
P.~Claus, M.~Gunaydin, R.~Kallosh, J.Rahmfeld and Y.~Zunger, Supertwistor as quarks of $SU(2,2|4)$, JHEP \textbf{9905} (1999) 019, \href{http://arxiv.org/abs/hep-th/9905112}{arXiv:hep-th/9905112}.
\bibitem{GMZ2}
M.~Gunaydin, D.~Minic and M.~Zagermann, $4d$ doubleton conformal theories, CPT and IIB string on $AdS_5\times S^5$, Nucl. Phys. \textbf{B534} (1998) 96, \href{http://arxiv.org/abs/hep-th/9806042}{arXiv:hep-th/9806042}.

\bibitem{MCP}
R.~Penrose and M.A.H.~MacCallum, Twistor theory: an approach to quantisation of fields and space-time, Phys. Rept. \textbf{6C} (1972) 241.

\bibitem{FT'85}
E.S.~Fradkin and A.A.~Tseytlin, Conformal supergravity, Phys. Rept. \textbf{119} (1985) 233.

\bibitem{Liu}
H.~Liu and A.A.~Tseytlin, $D=4$ Super Yang-Mills, $D=5$ gauged supergravity, and $D=4$ conformal supergravity, Nucl. Phys. \textbf{B533} (1998) 88, \href{http://arxiv.org/abs/hep-th/9804083}{arXiv:hep-th/9804083}.
\bibitem{Balasubramanian}
V.~Balasubramanian, E.~Gimon, D.~Minic and J.~Rahmfeld, Four dimensional conformal supergravity from $AdS$ space,
\href{http://arxiv.org/abs/hep-th/0007211}{arXiv:hep-th/0007211}.
\bibitem{Metsaev08}
R.R.~Metsaev, Shadows, currents and $AdS$, Phys. Rev. \textbf{D78} (2008) 106010, \href{http://arxiv.org/abs/0805.3472}{arXiv:0805.3472}.
\bibitem{Metsaev13}
R.R.~Metsaev, CFT adapted approach to massless fermionic fields, $AdS/CFT$, and fermionic conformal fields, \href{http://arxiv.org/abs/1311.7350}{arXiv:1311.7350}.
\bibitem{Eastwood-AMS}
M.G.~Eastwood, Supersymmetry, twistors, and the Yang-Mills equations, Trans. AMS \textbf{301} (1987) 615.
\bibitem{Mason}
L.~Mason and D.~Skinner, Ambitwistor strings and the scattering equation, JHEP {\bf 1407} (2014) 048, \href{http://arxiv.org/abs/1311.2564}{arXiv:1311.2564}.


\bibitem{U'16}
D.V.~Uvarov, Ambitwistors, oscillators and massless fields on $AdS_5$, Phys. Lett. \textbf{B762} (2016) 415, \href{http://arxiv.org/abs/1607.05233}{arXiv:1607.05233 [hep-th]}.

\bibitem{Geyer}
Y.~Geyer, A.E.~Lipstein and L.~Mason, Ambitwistor strings in four dimensions, Phys. Rev. Lett. \textbf{113} (2014) 081602, \href{http://arxiv.org/abs/1404.6219}{arXiv:1404.6219 [hep-th]}.
\bibitem{Tourkine}
E.~Casali and P.~Tourkine, On the null origin of the ambitwistor string, JHEP {\bf 1611} (2016) 036, \href{http://arxiv.org/abs/1606.05636}{arXiv:1606.05636 [hep-th]}.
\bibitem{U'17}
D.V.~Uvarov, Massless spinning particle and null-string on $AdS_d$: projective-space approach, J. Phys. A: Math. Theor. \textbf{51} (2018) 285402, \href{http://arxiv.org/abs/1707.05761}{arXiv:1707.05761 [hep-th]}.

\bibitem{BMS}
R.~Boels, L.~Mason and D.~Skinner, Supersymmetric gauge theories in twistor space, JHEP {\bf 0702} (2007) 014, \href{http://arxiv.org/abs/hep-th/0604040}{arXiv:hep-th/0604040}.

\bibitem{Wess}
J.~Wess and J.~Bagger, Supersymmetry and Supergravity, Princeton Univ. Press, 1992.

\end{thebibliography}
\end{document}